\documentclass[aps,amssymb,amsmath,prd,twocolumn,
showpacs,preprintnumbers,superscriptaddress,nofootinbib,floatfix]{revtex4-1}
\usepackage{graphicx}
\usepackage{bm}
\usepackage{mathrsfs}
\usepackage{latexsym}
\usepackage{float}
\usepackage{color}
\usepackage[normalem]{ulem} 
\usepackage{dcolumn}
\usepackage[colorlinks=true,citecolor=blue,urlcolor=blue]{hyperref}
\usepackage[usenames,dvipsnames]{xcolor}
\usepackage[caption=false]{subfig}
\usepackage{slashed}
\usepackage{orcidlink} 

\newcommand\Eqn[1]{Eq.~(\ref{#1})}  

\newcommand{\be}{\begin{equation}}
\newcommand{\ee}{\end{equation}}
\newcommand{\beq}{\begin{eqnarray}}
\newcommand{\eeq}{\end{eqnarray}}

\newcommand{\nb}{n_{\rm B}}

\newcommand{\Msolar}{{\rm M}_{\odot}}

\newcommand{\ep}{\varepsilon}
\newcommand{\eps}{\epsilon}

\newcommand{\csqeq}{c_{\rm eq}^2}
\newcommand{\csqad}{c_{\rm ad}^2}
\newcommand{\csqdy}{c_{\rm dy}^2}

\newcommand{\brv}{Brunt--V\"ais\"al\"a}
\newcommand{\gm}{$g$-mode~}
\newcommand{\gms}{$g$-modes~}
\newcommand{\fmo}{$f$-mode~}

\begin{document}
\preprint{N3AS-25-007, INT-PUB-25-011}
\title{Suppression of composition $g$-modes in chemically-equilibrating warm neutron stars} 

\author{Tianqi Zhao
\orcidlink{0000-0003-4704-0109}}
\email{tianqi.zhao@berkeley.edu}
\affiliation{Department of Physics and Astronomy, Ohio University,
Athens, OH~45701, USA}
\affiliation{Institute for Nuclear Theory, University of Washington, Seattle, WA~98195, USA}
\affiliation{Network for Neutrinos, Nuclear Astrophysics, and Symmetries (N3AS), University of California, Berkeley, Berkeley, CA~94720, USA}

\author{Peter~B. Rau
\orcidlink{0000-0001-5220-9277}}
\email{peter.rau@columbia.edu}
\affiliation{Columbia Astrophysics Laboratory, Columbia University, New York, NY~10027, USA}
\affiliation{Institute for Nuclear Theory, University of Washington, Seattle, WA~98195, USA}

\author{Alexander Haber
\orcidlink{0000-0002-5511-9565}}
\email{ahaber@physics.wustl.edu}
\affiliation{Mathematical Sciences and STAG Research Centre, University of Southampton, Southampton SO17 1BJ, United Kingdom}
\affiliation{Department of Physics, Washington University in St.~Louis, St.~Louis, MO~63130, USA}

\author{Steven~P. Harris
\orcidlink{0000-0002-0809-983X}}
\email{stharr@iu.edu}
\affiliation{Center for the Exploration of Energy and Matter and Department of Physics,
Indiana University, Bloomington, IN 47405, USA}

\author{Constantinos Constantinou
\orcidlink{0000-0002-4932-0879}}
\email{cconstantinou@ectstar.eu}
\affiliation{INFN-TIFPA, Trento Institute of Fundamental Physics and Applications, Povo, 38123 TN, Italy}
\affiliation{European Centre for Theoretical Studies in Nuclear Physics and Related Areas, Villazzano, 38123 TN, Italy}

\author{Sophia~Han
\orcidlink{0000-0002-9176-4617}}
\email{sjhan@sjtu.edu.cn}
\affiliation{Tsung-Dao Lee Institute, Shanghai Jiao Tong University, Shanghai~201210, China}
\affiliation{School of Physics and Astronomy, Shanghai Jiao Tong University, Shanghai~200240, China}
\affiliation{Network for Neutrinos, Nuclear Astrophysics, and Symmetries (N3AS), University of California, Berkeley, Berkeley, CA~94720, USA}
\affiliation{State Key Laboratory of Dark Matter Physics, Shanghai Jiao Tong University, Shanghai 201210, China}

\date{\today}

\begin{abstract}

We investigate the impact of chemical equilibration and the resulting bulk viscosity on non-radial oscillation modes of warm neutron stars at temperatures up to $T\approx 5$ MeV, relevant for protoneutron stars and neutron-star post-merger remnants. In this regime, the relaxation rate of weak interactions becomes comparable to the characteristic frequencies of composition $g$-modes in the core, resulting in resonant damping. To capture this effect, we introduce the dynamical sound speed, a complex, frequency-dependent generalization of the adiabatic sound speed that encodes both the restoring force and the dissipative effects of bulk compression. Using realistic weak reaction rates and three representative equations of state, we compute the complex frequencies of composition $g$-modes with finite-temperature profiles. We find that bulk viscous damping becomes increasingly significant with temperature and can completely suppress composition $g$-modes. In contrast, the $f$-mode remains largely unaffected by bulk viscosity due to its nearly divergence-free character. Our results highlight the sensitivity of $g$-mode behavior to thermal structure, weak reaction rates, and the equation of state, and establish the dynamical sound speed as a valuable descriptor characterizing oscillation properties in dissipative neutron star matter.

\end{abstract}


\maketitle
\section{Introduction}
\label{sec:intro}

Asteroseismology is a key probe of stellar interiors, and when applied to neutron stars (NSs) can be used to provide information on their chemical composition and the dense matter equation of state (EOS)~\citep{Andersson:2024amk,Andersson:2019yve}. Chemical composition is most sensitively probed by the $g$-modes, which are supported by gravity and a chemical composition gradient-driven buoyancy. These oscillation modes have frequencies in the tens to hundreds of Hz range. There is also an entropy gradient contribution to the buoyancy, but at typical temperatures for most of a neutron star's life ($T\lesssim \text{a few}$ {times} $10^8$ K), the chemical composition gradient is the dominant contribution to the $g$-mode, and purely thermal \gms have frequencies of order a few Hz~\citep{McD83} at these temperatures. 
The gradients of various particle species have been considered as the source of neutron star $g$-modes, starting with the proton fraction of total baryons~\citep{RG92}. Later, \gms due to the muon fraction of total leptons (usually in superfluid neutron stars)~\citep{Kantor:2014lja,Dommes:2015wul,Passamonti:2015oia,Yu:2016ltf,Rau2018}, hyperons~\citep{Dommes:2015wul,Tran:2022dva}, and quarks in the inner core~\citep{Wei:2018tts,Constantinou:2021hba,Zhao:2022toc,Kumar:2021hzo} were studied.

Recently, large modifications to \gms at $T>10$ MeV ($\gtrsim 10^{11}$ K) have been shown due to the effect of finite temperature on the EOS~\citep{Lozano:2022qsm}. 
At these temperatures, thermal contributions to \gms are comparable to the composition gradient-driven buoyancy~\citep{Ferrari2003,Gittins:2024oeh}. 
However, most calculations of NS \gms that focused on the chemical composition-driven modes have assumed cold (zero-temperature) neutron stars. This is a reasonable assumption for most of a neutron star's lifetime, since thermal effects on the mode spectrum for $T\lesssim 100$ keV ($\lesssim 10^9$ K) are generally very small and neutron stars cool below this within a few years after their birth~\citep{Page:2004fy}. The aforementioned references, including those that have included finite-temperature effects, thus typically assumed infinitely slow chemical reaction rates, a good approximation for cold neutron stars as long as the weak reactions in standard nuclear matter responsible for restoring beta equilibrium in perturbed fluid elements are much slower than the oscillation frequency at these temperatures. 

However, at temperatures in the MeV ($\sim10^{10}$ K) range, reaction rates are comparable to \gm oscillation frequencies. These temperatures are unlikely to be reached in inspiraling binary neutron stars~\citep{Counsell:2024pua}, where \gms may be detectable~\citep{Kwon:2024zyg}, since tidal excitation of oscillation modes are only expected to heat the stars to a few {times} $10^8$ K prior to merger~\citep{Lai:1993di,Arras:2018fxj}. MeV-range temperatures are thus relevant at two stages of a neutron star's lifetime: i) in young stars immediately after their birth in core-collapse supernova, and ii) in post-merger remnant neutron stars formed in binary mergers.
For instance, the \gms identified in numerical simulations as contributing to the gravitational-wave signal from protoneutron stars produced by core-collapse supernovae~\citep{Torres-Forne:2017xhv,Morozova:2018glm,Vartanyan:2023sxm,Wolfe:2023rfx,Jakobus:2023fru} will be modified by finite nuclear reaction rates, which was not considered in previous studies. 
These reactions 
give rise to bulk viscosity~\citep{Harris:2024evy}, which has impacts across neutron star physics, including in damping radial oscillation modes~\citep{Gusakov:2005dz}, $r$-modes~\citep{Andersson:1998ze}, and as an energy dissipation mechanism in neutron star mergers~\citep{Alford:2017rxf}. 
Bulk viscosity can be accounted for by including, in a reaction network, the underlying weak interaction rates (e.g., in radially oscillating neutron stars~\citep{Gusakov:2005dz}, in neutron stars at the edge of stability~\citep{Gourgoulhon:1993,Gourgoulhon:1995}, for inspiral neutron stars~\citep{Arras:2018fxj} or in neutron star merger simulations~\citep{Most:2022yhe,Espino:2023dei}), or through direct calculation of the bulk viscosity coefficient and implementation of dissipative relativistic hydrodynamics (e.g., in radially oscillating neutron stars~\citep{Camelio:2022ljs,Camelio:2022fds} or in neutron star merger simulations~\citep{Chabanov:2023blf}).

The main modification to the dynamics of neutron star oscillations due to nuclear reactions is to damp the composition $g$-modes. In an early investigation, 
\citet{Andersson:2019mxp} performed a local plane-wave analysis of the \brv~frequency, showing how arbitrary nuclear reaction rates can introduce a complex component to the \gm frequency, effectively damping the oscillations. Building on this,~\citet{Counsell:2023pqp} extended the study by computing global \gms in cold neutron stars using the BSk21 EOS. They showed that the real part of the \gm frequency decreases while the imaginary part increases as the reaction rate is increased.  Here, also, the \gms properties were calculated as a function of the nuclear reaction rate - disconnected from its actual value in dense matter. In realistic NS matter, these rates depend sensitively on the local temperature and composition profiles, and should be computed self-consistently once a specific equation of state is assumed.

To address this, we compute the \gms of hot neutron stars, incorporating microphysically-calculated temperature-dependent nuclear reaction rates consistent with the underlying EOS we use to compute the background stellar models and the oscillation modes. The physically grounded reaction rates lead to well-defined damping through bulk viscosity, moving beyond parametrized viscosity as in previous studies. By linking realistic microphysical inputs to global mode calculations, our approach enables a more accurate and self-consistent assessment of bulk viscous damping mechanisms in NSs under various thermal conditions. These rates have also been applied to examine the role of bulk viscosity in neutron star mergers through analyses of local density oscillations~\citep{Alford:2019kdw, Alford:2019qtm,Alford:2023gxq}, and numerical simulations~\citep{Most:2022yhe,Chabanov:2023blf}.  To clarify, we only consider the chemical composition gradient \gms at finite temperatures, and not the thermal \gm contribution, since while we consider $\sim$ MeV temperatures, we do not consider the $T\gtrsim 10$ MeV temperatures at which the thermal contribution to \gms is comparable to the chemical composition. Our quasinormal mode calculations use the zero-temperature EOS. {Since the EOS and the susceptibilities required to calculate the dynamic speed of sound show a very weak temperature dependence in the temperature range studied here, we use the $T=0$ approximation for those quantities, as justified in Appendix \ref{app:thermodynamics}.} Only the calculation of the weak interaction rates {and the calculation of the constant-entropy temperature profiles} include (as they must) temperature. We also ignore the effects of nucleonic superfluidity-superconductivity, assuming that Cooper pairing between nucleons is destroyed at the temperatures of interest, which are generally well above the critical temperatures for the $p$-wave paired neutron superfluidity ($T_{cn}\sim 10^9$ K) and $s$-wave paired proton superconductivity ($T_{cp}\sim 7\times10^9$ K) expected to be found in neutron star cores~\citep{Sedrakian:2018ydt}.

In Section~\ref{sec:framework} we discuss the origin of bulk viscosity in neutron stars and introduce the concept of the dynamical sound speed, which directly encodes the effects of bulk viscosity. This formalism can be naturally incorporated into the eigenvalue problem for calculating \gm frequencies in hot neutron stars, which we summarize in Section~\ref{sec:calcs}. We discuss the resulting changes to the $g$- and $f$-mode spectra of hot neutron stars due to bulk viscosity in Section~\ref{sec:results}, considering both isothermal and isentropic temperature profiles. Our conclusions and possible extensions of this work are presented in Section~\ref{sec:cons}. We work in $c=\hbar=1$ units.

\section{Bulk viscosity and dynamical sound speed}
\label{sec:framework}

Finite temperature, neutrinoless $npe$ matter out of chemical equilibrium is described by specifying the pressure $p(\nb,{\delta\mu},T)$, where $\nb$ is the baryon number density, $T$ is the temperature, and the degree to which the matter is out of beta equilibrium (used interchangeably with ``chemical equilibrium" in this work), is
\begin{equation}
    {\delta\mu} \equiv \mu_n-\mu_p-\mu_e.
\end{equation}
At sufficiently high temperatures, the neutrino mean free path shrinks~\citep{Alford:2018lhf,Roberts:2016mwj} and neutrinos become trapped and the definition of beta equilibrium is modified~\citep{Alford:2019kdw,Raduta:2020fdn}.  Neutrino-trapping effects are likely present at the higher end of the temperature range we consider here, but we neglect neutrino trapping, for simplicity. 
The effect of neutrino-trapping on compositional \gm frequencies has not been studied. 
Instead of specifying the temperature, it is sometimes convenient to instead specify the entropy per baryon $S\equiv s/\nb$, so that the pressure becomes the function $p(\nb,{\delta\mu},S)$. 
A local, adiabatic, perturbation to the pressure $p(\nb,{\delta\mu},S)$ is given by 
\begin{eqnarray}
\Delta p &=&  \left.\dfrac{\partial p}{\partial \nb}\right|_{{\delta\mu},  S}\Delta \nb + \left.\dfrac{\partial p}{\partial {\delta\mu}}\right|_{\nb,  S}\Delta {\delta\mu}.
\label{eq:perturbation_p}
\end{eqnarray}
The proton fraction, $x\equiv n_p/\nb$, in beta equilibrium becomes a function of density and temperature $x_{\rm eq}(\nb, T)$.  This can be calculated for various EOSs.  When the proton fraction is away from its beta equilibrium value by an amount 
$\delta x=x-x_{\rm eq}$, the system is out of chemical equilibrium by an amount ${\delta\mu}$. 
The perturbation of ${\delta\mu} (\nb,\delta x,S)$ is thus
\begin{eqnarray}
\Delta{\delta\mu} &=&  \left.\dfrac{\partial {\delta\mu}}{\partial \nb} \right|_{\delta x,S} \Delta \nb + \left.\dfrac{\partial {\delta\mu}}{\partial \delta x} \right|_{\nb,S} \Delta \delta x.
\label{eq:perturbation_mu}
\end{eqnarray}
At fixed density, a fluid element of $npe$ matter that is out of beta equilibrium by a small amount ${\delta\mu}$ adjusts its proton fraction back to its beta equilibrium value accordingly, 
to first order in ${\delta\mu}$, 
\begin{equation}
\dfrac{\partial \delta x}{\partial t}=\dfrac{\lambda_{\rm{n}\leftrightarrow\rm{p}}}{\nb}{\delta\mu} \, ,
\label{eq:bulk_linear}
\end{equation}
where 
\begin{equation}
\lambda_{\rm{n}\leftrightarrow\rm{p}} (T,\nb) \equiv  \dfrac{\partial(\Gamma_{n\rightarrow p}-\Gamma_{p\rightarrow n})}{\partial {\delta\mu}}\bigg\vert_{{\delta\mu}=0}.\label{eq:net_decay_rate}
\end{equation}
The decay rates (number of decays per volume per time) $\Gamma_{n\rightarrow p}$ and $\Gamma_{p\rightarrow n}$ are of the flavor-changing processes that inter-convert neutrons and protons, which in $npe$ matter are the Urca processes, namely neutron decay and electron capture~\citep{Yakovlev:2000jp}.  For small deviations from beta equilibrium, taking the time derivative of \Eqn{eq:perturbation_mu} and then plugging in \Eqn{eq:bulk_linear}, yields 
\begin{equation}
\dfrac{\partial\Delta{\delta\mu}}{\partial t} =  \left.\dfrac{\partial {\delta\mu}}{\partial \nb} \right|_{\delta x,S}\dfrac{\partial\Delta \nb}{\partial t} -\gamma \Delta{\delta\mu}, \label{eq:perturbation_dtmu}
\end{equation} 
where we have recast $\lambda_{n\leftrightarrow p}$ as the beta equilibration relaxation rate $\gamma$, which is defined as 
\begin{equation}
\gamma  \equiv -\dfrac{\lambda_{\rm{n}\leftrightarrow\rm{p}}}{\nb}\left.\dfrac{\partial{\delta\mu}}{\partial \delta x}\right|_{\nb,S},\label{eq:gamma_def}
\end{equation} 
which has a dimension of frequency and can be directly compared to the oscillation mode angular frequency $\omega$. The calculation of the susceptibility $(\partial{\delta\mu}/\partial \delta x)\vert_{\nb,S}$ from the EOS tables, using $\nb$, $x$ and $T$ as input variables, is detailed in Appendix~\ref{app:thermodynamics}.

\begin{figure}
\centering
\includegraphics[width=\linewidth]{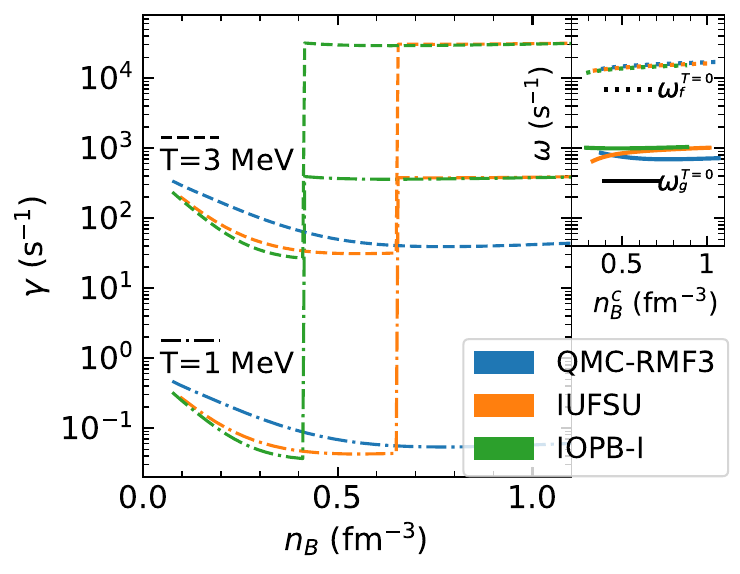}
\caption{Left (main) panel 
shows the beta equilibration relaxation rate $\gamma$ defined in \Eqn{eq:gamma_def} for QMC-RMF3, IUFSU and IOPB-I EOSs at temperatures $T=1$ MeV (dot-dashed) and $T=3$ MeV (dashed). 
The solid (dashed) lines in the top right panel represent the \gm ($f$-mode) frequency of cold NSs with various central baryon densities $\nb^c$. 
The sharp jump in the rate is due to the onset of the direct Urca threshold.}
\label{fig:response}
\end{figure}

The calculation of the rates in \Eqn{eq:gamma_def} requires a microscopic model of nuclear matter that provides us not just with the EOS, but also with the dispersion relations of the nucleons. We calculate our results for three different EOSs, described below, in order to properly consider the range of possible \gm physics.   

In Fig.~\ref{fig:response}, we plot the beta equilibration rate $\gamma$ as a function of the baryon density. We include both the direct Urca processes ($n\rightarrow p+e^-+\bar{\nu}_e$ and $e^-+p\rightarrow n + \nu_e$) as well as the modified Urca processes ($n+N\rightarrow p+e^-+\bar{\nu}_e+N$ and $e^-+p+N\rightarrow n + \nu_e+N$, where $N$ is a neutron or proton). 
For simplicity, we calculate the rates of these processes in the Fermi surface approximation, which is valid for strongly degenerate nuclear matter (typically, $T\lesssim 1\text{ MeV}$)~\citep{Alford:2018lhf,Alford:2021ogv}. In this limit, only particles on the Fermi surface participate in the reactions. This constraint on the phase space gives rise to a direct Urca threshold density: below this density, direct Urca is kinematically forbidden because the neutron Fermi momentum is too big relative to the proton and electron Fermi momenta (equivalently, the proton fraction is too small).  Modified Urca does not have such a density threshold and has a rate that is only weakly dependent on the baryon density~\citep{Yakovlev:2000jp}. As the temperature rises, the beta relaxation rate $\gamma$ initially scales as $T^4$ for the direct Urca process and as $T^6$ for the modified Urca process. Above temperatures of a few MeV, thermal contributions from particles away from the Fermi surface begin to dominate the reaction rates, causing a blurring in the density of the direct Urca threshold~\citep{Alford:2018lhf,Alford:2021ogv}. 
The formulas for the Fermi surface approximation of the Urca rates are given in App.~A and B of~\citet{Most:2022yhe}.

The rates shown in Fig.~\ref{fig:response} were calculated in matter described by three different EOSs, each built from a relativistic mean-field theory (RMFT), QMC-RMF3~\citep{Alford:2022bpp, Alford:2023rgp}, IUFSU~\citep{Fattoyev:2010mx}, and IOPB-I~\citep{Kumar:2017wqp}. All of them agree with current astrophysical and theoretical constraints on the EOS~\citep{MUSES:2023hyz}, but differ in their predictions of the composition of $npe$ matter, which leads to significantly different predictions of the Urca rates.  Critically, they have different predictions for the direct Urca threshold. 
While QMC-RMF3 does not have a direct Urca threshold at all, IOPB-I has a threshold density of $n_{\rm{B},\mathrm{thr.}}=0.414~\mathrm{fm}^{-3}$, and IUFSU a threshold density of $n_{\rm{B},\mathrm{thr.}}=0.655~\mathrm{fm}^{-3}$. 

At low density, in Fig.~\ref{fig:response}, direct Urca is forbidden for all EOSs and the rates are entirely due to modified Urca.  As the density increases, the rates for two of the EOSs, IUFSU and IOPB-I, increase dramatically as the direct Urca threshold is passed. The QMC-RMF3 EOS always forbids direct Urca, as the proton fraction never rises above $1/9$ (c.f.~\citet{Lattimer:1991ib}). Equilibration rates at two different temperatures ($T = 1\text{ and } 3\text{ MeV}$) are shown: clearly, the equilibration rate is a strong function of temperature (see e.g., Fig.~8 in~\citet{Harris:2024evy}). To the right of the main plot, we attached a plot of the oscillation frequencies of the \fmo and \gm as a function of the \textit{central density} of the neutron star (different from the x-axis of the left plot).  This presentation allows us to compare the frequency of the density change of the matter $\omega$ with the rate at which beta equilibrium is restored $\gamma$.  This ratio is the key quantity in determining the bulk-viscous damping of the oscillation modes and the disappearance of the \gm at high temperature.  Evidently, temperatures of a few MeV will make the beta equilibration rate comparable to the density oscillation frequency.

Applying the harmonic oscillation ansatz ${\rm e}^{i\omega t}$ for all perturbations in \Eqn{eq:perturbation_dtmu}, we obtain 
\begin{eqnarray}
\Delta{\delta\mu} &=&  \dfrac{\left.\dfrac{\partial {\delta\mu}}{\partial \nb} \right|_{\delta x,S} \Delta \nb}{1+ \dfrac{\gamma}{i\omega}} \,.
\label{eq:harmonic_mu}
\end{eqnarray}
We substitute \Eqn{eq:harmonic_mu} into \Eqn{eq:perturbation_p}, yielding 
\begin{eqnarray} 
\Delta p &=& \dfrac{p\Gamma}{\nb} \Delta \nb \, ,\label{eq:perturbation_p_gamma} 
\end{eqnarray}
where the complex-valued effective adiabatic index for a damped oscillation of frequency $\omega$ is given by 
\begin{eqnarray} 
\Gamma &=&  \dfrac{\nb}{p} \left[\left.\dfrac{\partial p}{\partial \nb}\right|_{{\delta\mu},  S}+\dfrac{\left.\dfrac{\partial p}{\partial {\delta\mu}}\right|_{\nb,  S} \left.\dfrac{\partial {\delta\mu}}{\partial \nb} \right|_{\delta x,S} }{1+ \dfrac{\gamma}{i\omega}}\right]. 
\end{eqnarray}
This expression can be recast in a more transparent form: 
\begin{eqnarray}
\Gamma &=&  \Gamma_{\rm eq}+\dfrac{\Gamma_{\rm ad}-\Gamma_{\rm eq}}{1+ \dfrac{\gamma}{i\omega}} 
\label{eq:GammaComplex}\, ,\\
\Gamma_{\rm ad}&=&  \dfrac{\nb}{p} \left.\dfrac{\partial p}{\partial \nb}\right|_{\delta x,  S}
\label{eq:gamma_ad}\, ,\\
\Gamma_{\rm eq}&=&  \dfrac{\nb}{p} \left.\dfrac{\partial p}{\partial \nb}\right|_{{\delta\mu},  S} \, ,
\label{eq:gamma_eq}
\end{eqnarray}
where $\Gamma_{\rm ad}$ and $\Gamma_{\rm eq}$ characterize the compressibility of matter in the slow and fast limits of beta equilibration, respectively. The computation of these derivatives from the EOS tables, using $\nb$, $x$ and $T$ as input variables, is described in detail in Appendix~\ref{app:thermodynamics}.

\begin{figure}
    \centering
    \includegraphics[width=\linewidth]{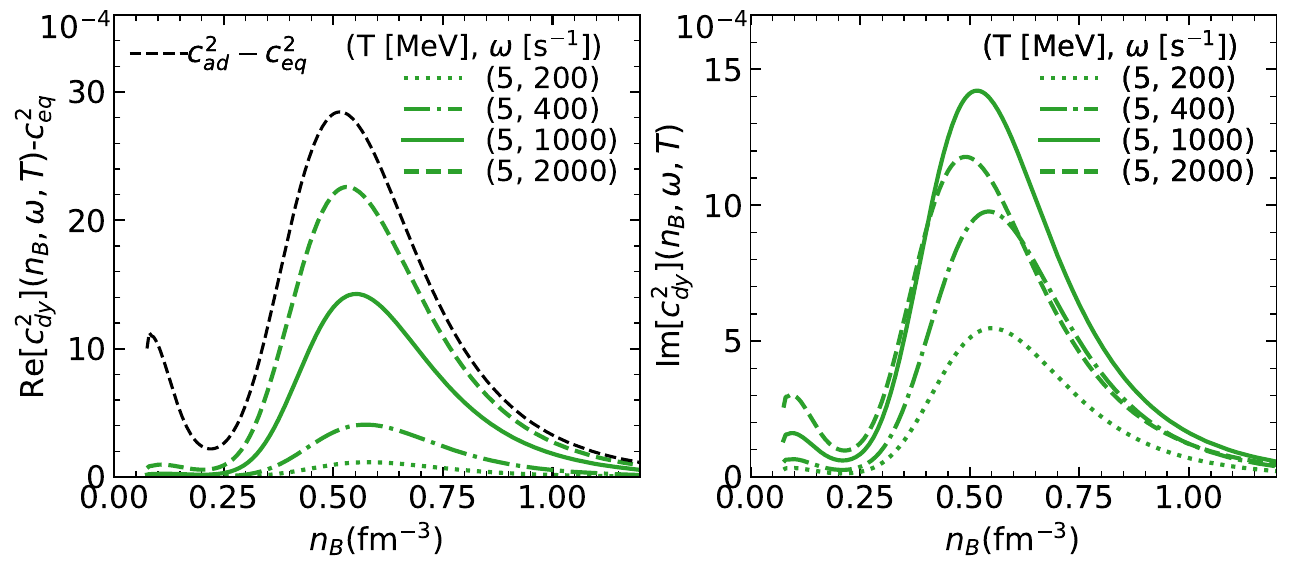}
    \includegraphics[width=\linewidth]{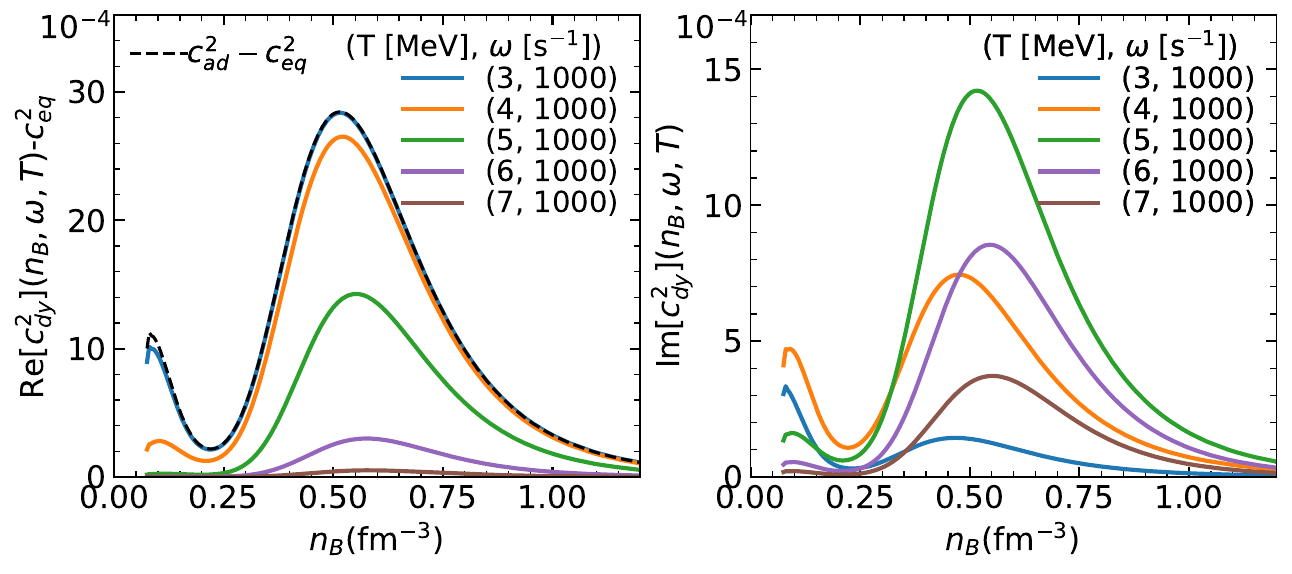}
\caption{Real (left panels) and imaginary (right panels) part of the dynamical sound speed squared for the QMC-RMF3 EOS. For the real part, the equilibrium sound speed squared is used as a reference. The black dashed line represents the difference between adiabatic and equilibrium squared sound speeds. The upper two panels correspond to fixed temperature $T=5$ MeV while varying frequency 
{$\omega= [200,400,1000,2000]$ s$^{-1}$}; the bottom two panels represent fixed frequency 
{$\omega=1000$ s$^{-1}$}
while varying temperature from 
{3 to 7} MeV. Note the different scales of y-axes in real and imaginary parts. 
}
\label{fig:cs2dy}
\end{figure}
\begin{figure}
\centering
\includegraphics[width=\linewidth]{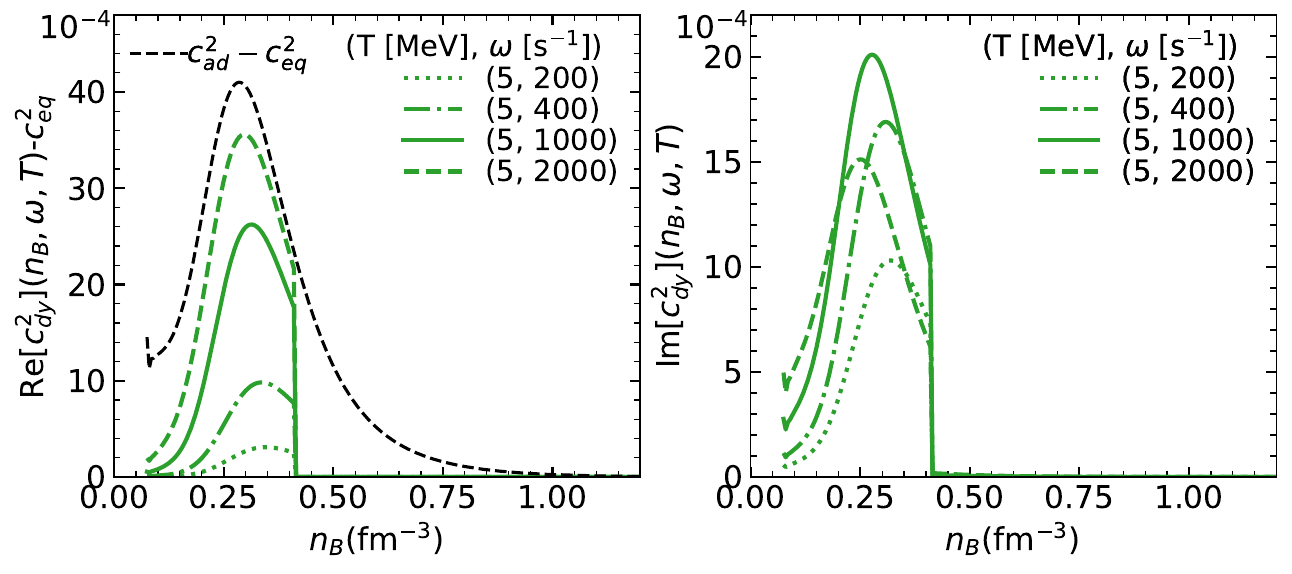}
    \includegraphics[width=\linewidth]{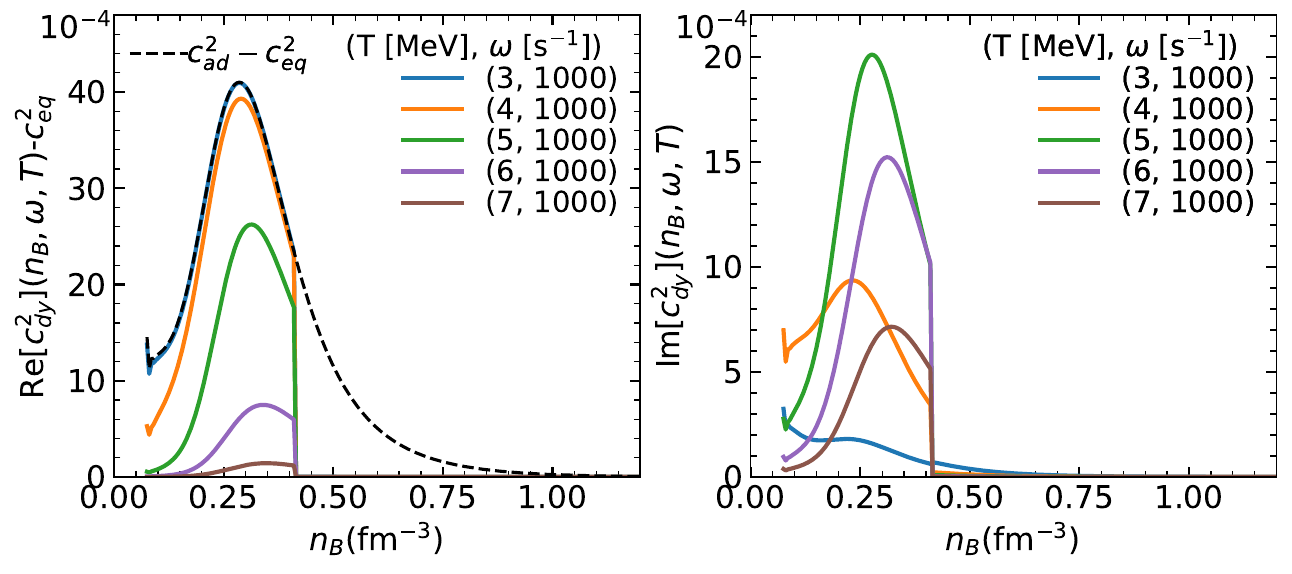}
\caption{Same as Fig.~\ref{fig:cs2dy}, but for the IOPB-I EOS.}
\label{fig:cs2dy_dUrca}
\end{figure}
The dynamical sound speed squared $\csqdy$, which incorporates finite-temperature viscous effects, is defined as
\begin{eqnarray}
\csqdy &=& \dfrac{p}{\ep+p}\,\Gamma = \csqeq +\dfrac{\csqad-\csqeq}{1+\dfrac{\gamma}{i\omega}} \, ,\label{eq:cs2dy}
\end{eqnarray}
where the equilibrium and adiabatic sound speed squared $\csqeq$ and $\csqdy$ are given by
\begin{equation}
\csqeq=\dfrac{p}{\ep+p}\,\Gamma_{\rm eq}\, , \qquad \csqad=\dfrac{p}{\ep+p}\,\Gamma_{\rm ad}\, . \label{eq:cs2ad_cs2eq}
\end{equation}
The real and imaginary parts of the dynamical sound speed squared are then given by
\begin{eqnarray}
{\rm Re}\left[\csqdy\right] &=& \csqeq+\left(\csqad-\csqeq\right)\dfrac{\omega^2}{\omega^2+\gamma^2} \, ,\label{eq:csq_dy_real}\\
{\rm Im}\left[\csqdy\right] &=& \left(\csqad-\csqeq\right)\dfrac{\omega\gamma}{\omega^2+\gamma^2}. \label{eq:csq_dy_imag}
\end{eqnarray}
The imaginary part, $ {\rm Im}\left[\csqdy\right]$, is directly related to the bulk viscosity coefficient $\zeta$, which is defined via the energy dissipation rate  $\mathop{{\rm d}\ep}/\mathop{{\rm d}t}=-\zeta(\nabla\cdot\mathbf{v})^2$, since the bulk viscosity coefficient is also equal to the EOS-related factor times the resonance expression $\gamma/\left(\gamma^2+\omega^2\right)$~\citep{Harris:2024evy}. Using thermodynamic identities and the definitions of the adiabatic indices [Eqs.~(\ref{eq:gamma_ad}) and~(\ref{eq:gamma_eq})], the bulk viscosity can be expressed as

\begin{eqnarray}
\zeta &=&\dfrac{\ep+p}{\omega}{\rm Im}\left[\csqdy\right]\\
&=& 
{4.5\times 10^{29} \rm{~g~cm}^{-1}\rm{~s}^{-1} \left(\dfrac{\omega}{600 \rm{~s^{-1}}}\right)^{-1}} \, \nonumber\\
&& \times \left(\dfrac{(\ep+P)/c^2}{3\times 10^{14}\rm{~g~ cm}^{-3}}\right)\left(\dfrac{{\rm Im}\left[\csqdy\right]}{0.001}\right).
\label{eq:bv_numerical}
\end{eqnarray}

For neutron star matter, the adiabatic sound speed is usually very close to the equilibrium sound speed ($\csqad-\csqeq \lesssim 0.01$), and therefore the second term on the right-hand side of \Eqn{eq:cs2dy} is small. As a result, the imaginary part of the dynamical sound speed squared is small, while the real part lies between the adiabatic and equilibrium squared sound speeds. The source of the difference between $\csqad$ and $\csqeq$ in our paper is the gradient in the proton fraction $x$, not a gradient in the entropy per particle $S$, {since we ignore finite temperature effects in the EOS as justified in Appendix \ref{app:thermodynamics}.} 
To demonstrate the dynamical sound speed squared $\csqdy$, we plot the real and imaginary part of $\csqdy$ versus $\nb$ for various temperatures $T$ and frequencies $\omega$ for the QMC-RMF3 EOS in Fig.~\ref{fig:cs2dy} and for the IOPB-I EOS in Fig.~\ref{fig:cs2dy_dUrca}. QMC-RMF3 shows double peaks in $\csqad-\csqeq$ due to a plateau in the equilibrium composition $x_{\rm eq}$ around $\nb=0.2$ fm$^{-3}$. The IOPB-I curve only has one peak and terminates at $\nb=0.414$ fm$^{-3}$, beyond which $\csqdy\approx\csqeq$ due to the onset of the direct Urca process. 

The real part of the dynamical sound speed is the adiabatic sound speed at low temperatures (or in the high-frequency limit), and as temperature is increased (or frequency is lowered), it decreases to the equilibrium sound speed. 
The imaginary part of the dynamical sound speed is non-monotonic with respect to temperature -- it starts from zero at low temperatures and then increases until reaching a peak at the resonant temperature where $\gamma=\omega$. 
As the temperature (and thus $\gamma$) further increases, the imaginary part of the dynamical sound speed drops to zero.  This resonant behavior is, naturally, just like the resonant behavior of bulk viscosity~\citep{Harris:2024evy}. 

\begin{figure}
\centering
\includegraphics[width=\linewidth]{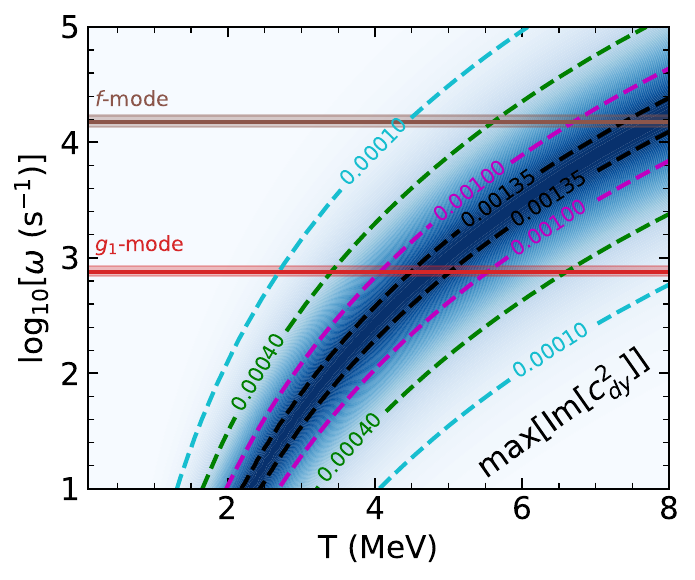}
\includegraphics[width=\linewidth]{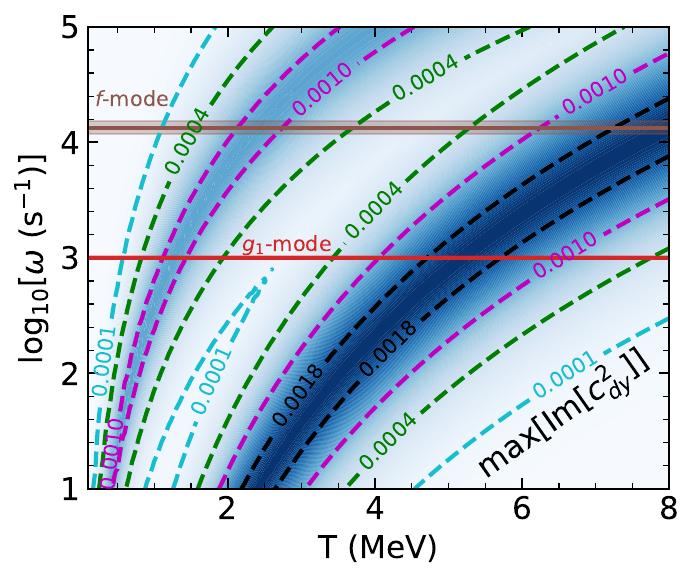}
\caption{
The dashed contour shows the maximum (across all densities) of the imaginary part of the dynamical sound speed squared for QMC-RMF3 (top) and IOPB-I (bottom). The dark blue region corresponds to the resonant peak, where bulk viscosity is most effective at damping oscillations. The horizontal brown and red bands represent typical \fmo and $g_1$-mode frequencies (at zero temperature) for NSs of various masses, with the central line corresponding to a 1.4 $\Msolar$ NS. 
}
\label{fig:contour_resonance}
\end{figure}

In Fig.~\ref{fig:contour_resonance}, we show the maximum value (across all densities) of the imaginary part of the dynamical sound speed squared in the temperature-frequency plane. This plot depicts the resonant behavior described above.  There is only one resonant peak for QMC-RMF3, while for IOPB-I, there exist two distinct resonant bands corresponding to the direct and modified Urca processes, respectively. These peaks are unrelated to the double-peak structure discussed in the context of Fig.~\ref{fig:cs2dy}.  The direct Urca process, with a higher reaction rate, has a resonant peak at a higher frequency or a lower temperature compared to the modified Urca process peak. Typical \fmo and \gm frequencies at zero temperature are shown in Fig.~\ref{fig:contour_resonance}.

In this work, the crust EOSs in beta equilibrium are constructed with the compressible liquid droplet model with fixed surface tension parameters $\sigma_s=1.2$ MeV fm$^{-2}$, $S_S=48$ MeV~\citep{Lattimer:1991nc}. Below the crust-core transition density, the adiabatic sound speed is set equal to the equilibrium sound speed, as our focus is on core \gms following~\citet{Jaikumar:2021jbw}. This is justified because the adiabatic sound speed in the crust has only a limited impact on the properties of core \gms\citep{Sun:2025zpj}.

\section{Mode equations}
\label{sec:calcs}

The components of the interior metric tensor for a spherically symmetric, non-rotating star are defined by
\begin{equation}
    {\rm d}s^2 = -{\rm e}^{\nu(r)}{\rm d}t^2+{\rm e}^{\lambda(r)}{\rm d}r^2+r^2({\rm d}\theta^2+\sin^2\theta{\rm d}\phi^2).
\end{equation}
Following e.g.,~\citet{Jaikumar:2021jbw}, we solve the general relativistic non-radial mode equations in the relativistic Cowling approximation, which ignores perturbations of the metric tensor. The mode equations are
\begin{align}
\dfrac{{\rm d}U}{{\rm d}r}={}&\dfrac{g}{\csqdy}U+{\rm e}^{\lambda/2}\left(\dfrac{l(l+1)}{\omega^2}{\rm e}^{\nu}-\dfrac{r^2}{\csqdy}\right)V,
\label{eq:dUdr}
\\
\dfrac{{\rm d}V}{{\rm d}r}={}&{\rm e}^{\lambda/2-\nu}\dfrac{\omega^2-N^2}{r^2}U+g\left(\dfrac{1}{\csqeq}-\dfrac{1}{\csqdy}\right)V,
\label{eq:dVdr}
\end{align}
where $U$ and $V$ are related to the radial component of the Lagrangian displacement field $\xi_r$ and the Eulerian perturbation of the pressure $\delta p$ by
\begin{equation}
    U = r^2{\rm e}^{\lambda/2}\xi_r, \qquad V=\dfrac{\delta p}{\ep+p}.
\end{equation}
$l$ is the degree of the vector spherical harmonic representing the angular dependence of the displacement field, which we restrict to $l=2$. The gravitational acceleration $g$ is 
\begin{equation}
g = -\dfrac{1}{\ep+p}\dfrac{{\rm d} p}{{\rm d} r}= \dfrac{1}{2}\dfrac{{\rm d}\nu}{{\rm d}r}.
\end{equation}
The two sound speeds squared $\csqeq$ and $\csqdy$ are given by \Eqn{eq:cs2dy} and \Eqn{eq:cs2ad_cs2eq}. Note that $\csqdy$ replaces $c_{\rm ad}^2$ from the original formulation of the mode equations, which are recovered in the limit that $\gamma\rightarrow 0$. Finally, the 
\brv~frequency squared $N^2$ is
\begin{equation}
N^2=g^2\left(\dfrac{1}{\csqeq}-\dfrac{1}{\csqdy}\right){\rm e}^{\nu-\lambda}.\label{eq:bv_frequency}
\end{equation}
Note that $N^2$ is now complex and depends on $\omega$ due to the inclusion of $\csqdy$. 
$N$ follows a similar temperature dependence as $\csqdy-\csqeq$ in Figs.~\ref{fig:cs2dy} and \ref{fig:cs2dy_dUrca}. Its real part decreases monotonically with increasing temperature, while the imaginary part starts from zero at zero temperature, increases to a maximum at $\gamma = \sqrt{3}\,\omega$, and then decreases with further temperature increase.

The boundary conditions at $r=0$ are
\begin{align}
U(r=0)=&\dfrac{l}{\omega^2}Y_0 r^{l+1},
\\
V(r=0)=&\dfrac{\ep}{\ep+p}Y_0 r^{l},
\end{align}
where $Y_0$ is a constant that shifts the overall amplitude of the oscillation mode, which is itself unconstrained. The $r=R$ boundary condition is the vanishing of the Lagrangian pressure perturbation at $\Delta p(r=R)=0$, which is equivalent to
\begin{equation}
0=V(r=R)-\dfrac{1}{2R^2}U(r=R){\rm e}^{-\lambda(r=R)/2}\left.\dfrac{{\rm d}\nu}{{\rm d}r}\right|_{r=R}.
\end{equation}
Since $c_{\rm dy}^2$ is complex, the solutions to the mode equations are also complex, therefore Eqs.~(\ref{eq:dUdr}) and (\ref{eq:dVdr}) can be solved by splitting them into real and imaginary parts. This is presented in Appendix~\ref{app:RealImagSplit}. We performed independent calculations of the oscillation modes using complex equations and then splitting into real and imaginary parts, obtaining identical results in each case.

To assess the validity of the relativistic Cowling approximation, we also carried out linearized full general relativity (GR) calculations of non-radial \fmo and $g$-modes, following the formalism outlined in the appendix of~\citet{Zhao:2022tcw}. The system of differential equations governing linear non-radial oscillations in full GR reduces to Eqs.~(\ref{eq:dUdr}) and (\ref{eq:dVdr}) of the relativistic Cowling approximation when the metric perturbations are neglected, as shown explicitly in~\citet{Zhao:2022toc}. {The main modification in this work is that we allow the adiabatic index or sound speed to be complex. We find the method of solving for eigenmodes based on root-finding of amplitude of ingoing gravitational wave component remains effective.}

\section{Results}
\label{sec:results}

\begin{figure}
\centering
\includegraphics[width=\linewidth]{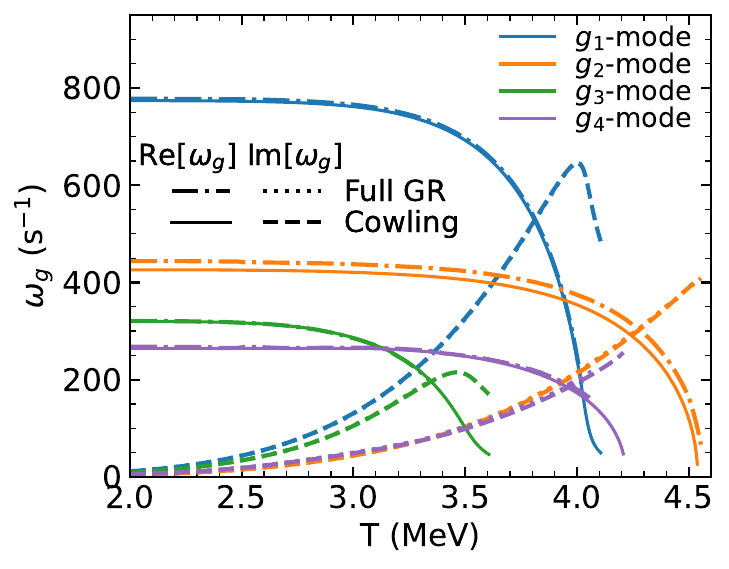}
\caption{Real and imaginary parts of the frequencies of the fundamental and the first three overtone \gms for a $1.4~\Msolar$ neutron star versus varying temperature for the QMC-RMF3 EOS. The imaginary part of the full GR calculation (dotted) largely overlaps with that of the calculation with Cowling approximation (dashed), and the real parts of the full GR calculation (dash-dotted) and Cowling approximation (solid) are also very similar. We define the ordering of \gms by their smooth asymptotic eigenmode at zero temperature.}
\label{fig:omega_g_vs_T}
\end{figure}
\begin{figure}
\subfloat[]{{\includegraphics[width=\linewidth]{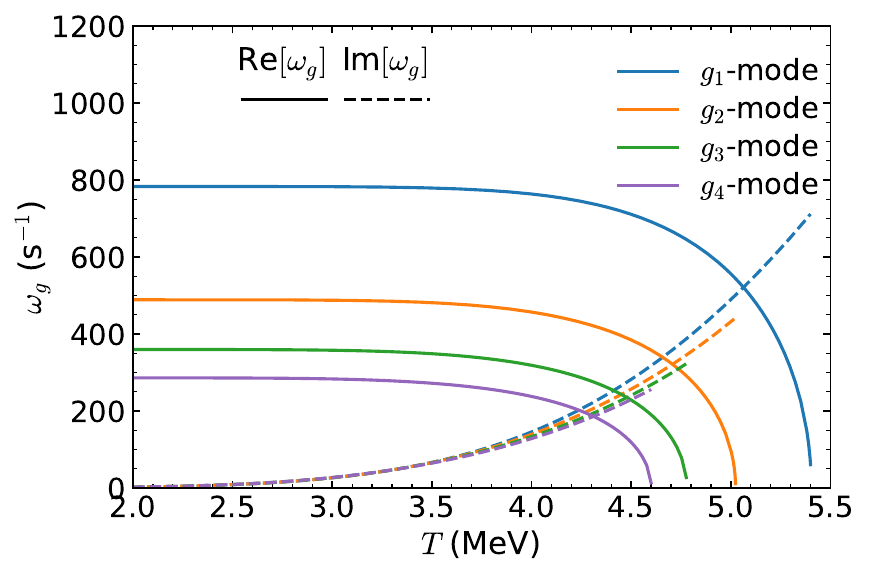}}}
\quad
\subfloat[]{{\includegraphics[width=\linewidth]{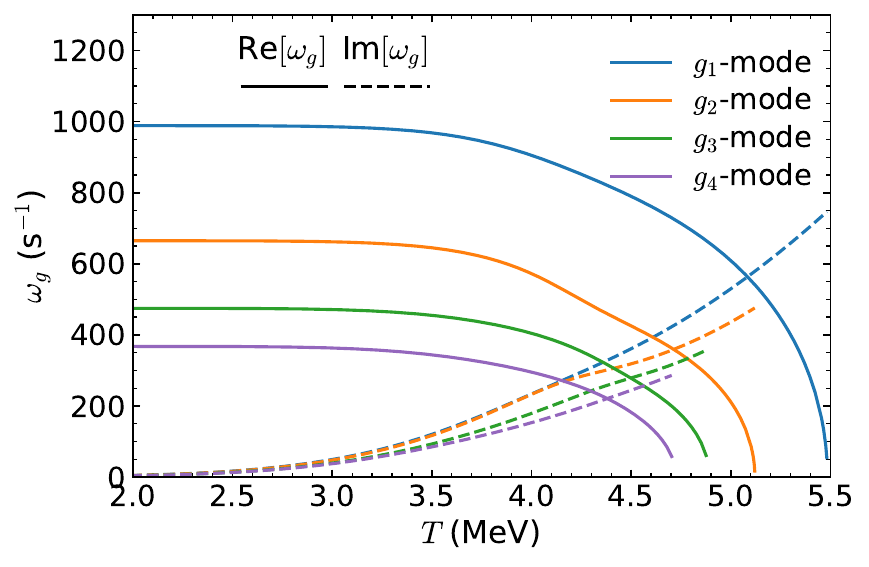}}}
\caption{Same as Fig.~\ref{fig:omega_g_vs_T} but for the IUFSU EOS (panel (a)) and the IOPB-I EOS (panel (b)).}
\label{fig:omega_g_vs_T_AltEOS}
\end{figure}

To examine the effect of weak interactions on the damping of $g$-modes, we computed the complex \gm frequencies of $1.4~\Msolar$ NSs at various temperatures for the three selected EOSs. 
{First, a note on mode classification. A standard way to classify the oscillation modes of a neutron star is the Cowling scheme, based on (a) the restoring force for the mode ($p$, $f$, $g$, etc.), which determines the typical frequency range; (b) the number of nodes of the mode's displacement field. We also follow this scheme to classify the $g$-modes, with a slight modification due to our inclusion of temperature-dependent bulk viscosity explicitly in the mode calculation. This modification is that
we base our classification on the mode behavior at $T=0$, and then continuously follow the same mode as the temperature is increased. For the complex frequency modes we study, the number of nodes can change as a function of temperature as the bulk viscous damping increases, the size of the imaginary part of the frequency increased and the real part decreases. However, we do not change the labeling of the mode as this happens: doing so results in a large number of changes of the mode labels at high temperature and is more confusing than clarifying. As long as we carefully follow each mode from zero temperature until the real part approaches zero, this classification is unambiguous. Since the mode frequencies are complex and well-separated in the complex plane, there are no avoided crossings between $g$-modes.}

Figure~\ref{fig:omega_g_vs_T} shows the real and imaginary parts of the \gm frequency as a function of temperature for the QMC-RMF3 EOS, with the results compared between using and not using the Cowling approximation. 
The fundamental $g$-mode, $g_1$, which has one radial node, and the lowest three overtone $g$-modes $\{g_2,g_3,g_4\}$, with two, three and four radial nodes respectively, are shown. 
The Cowling approximation introduces only a very small error compared to the full GR calculation for the $g$-modes. Therefore, for the remainder of the paper we only consider results obtained using the Cowling approximation.
As the temperature increases, the bulk viscosity rises until it reaches its resonant maximum around $T\approx5$ MeV, as can be read off from Fig.~\ref{fig:cs2dy}. With rising temperature, the real part of \gm frequency decreases while the imaginary part increases. This is an interplay between the bulk viscosity, that dampens the modes, and the faster relaxation rates. In the limit of 
instantaneous relaxation, the restoring buoyant force vanishes and thus the \gms vanish. 
This behavior is expected, as the \gm manifests as a global oscillation governed by the local \brv~frequency, 
\Eqn{eq:bv_frequency}. 
Since the beta relaxation rate $\gamma$ increases rapidly with temperature, scaling as $T^4$ for the direct Urca process and as $T^6$ for the modified Urca process, the dynamical sound speed reduces to the equilibrium sound speed at high temperature according to Eqs.~(\ref{eq:csq_dy_real})-(\ref{eq:csq_dy_imag}). 
Consequently, the \brv~frequency decreases with increasing temperature and ultimately vanishes.
The real parts of the 
odd overtone \gms 
(those modes with an odd number of radial nodes) vanish at lower temperatures compared to the 
even \gms because of the double peaks in the sound speed squared difference shown in Fig.~\ref{fig:cs2dy}. This leads to an interchange in the temperature at which these modes vanish, such that the 
$g_2$ and 
$g_4$-modes vanish at higher temperatures than the 
$g_1$ and 
$g_3$-modes. In comparison, Fig.~\ref{fig:omega_g_vs_T_AltEOS} shows the $g$-mode frequencies for the IUFSU and IOPB-I EOSs. Since these two EOSs have only one peak in their squared sound speed differences, they do not display a characteristic difference between the odd and even $g$-modes like that shown for the QMC-RMF3 EOS. 

\begin{figure}
\centering
\includegraphics[width=\linewidth]{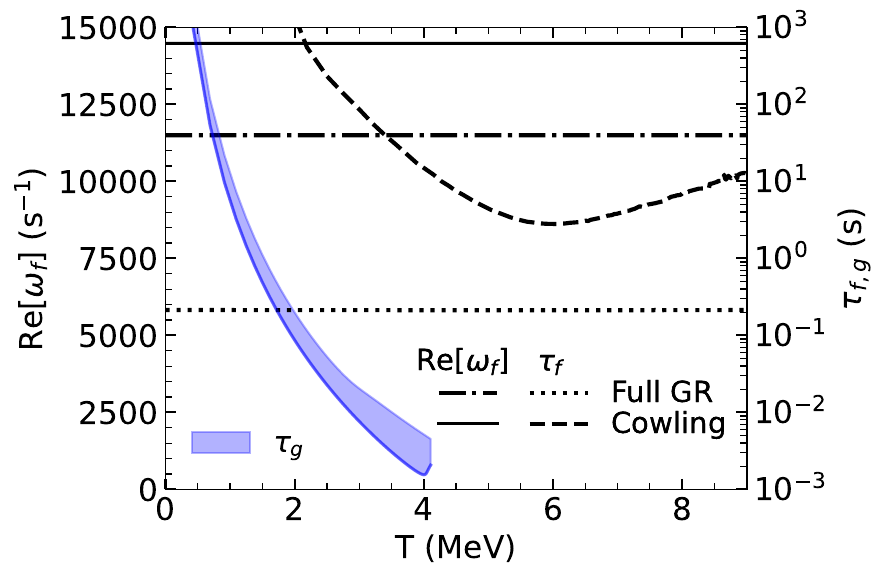}
\caption{Frequency (solid and dot-dashed) and damping time (dashed and dotted) of the \fmo calculated with and without Cowling approximation for the QMC-RMF3
EOS. 
The damping time of \gms in Fig.~\ref{fig:omega_g_vs_T} is plotted as the blue region for comparison. The bottom of the blue region corresponds to the lowest order $g$-mode.} 
\label{fig:damping_time_f}
\end{figure}
\begin{figure*}
\centering
\subfloat[QMC-RMF3, 
$g_1$-mode]{\includegraphics[width=0.38\linewidth]{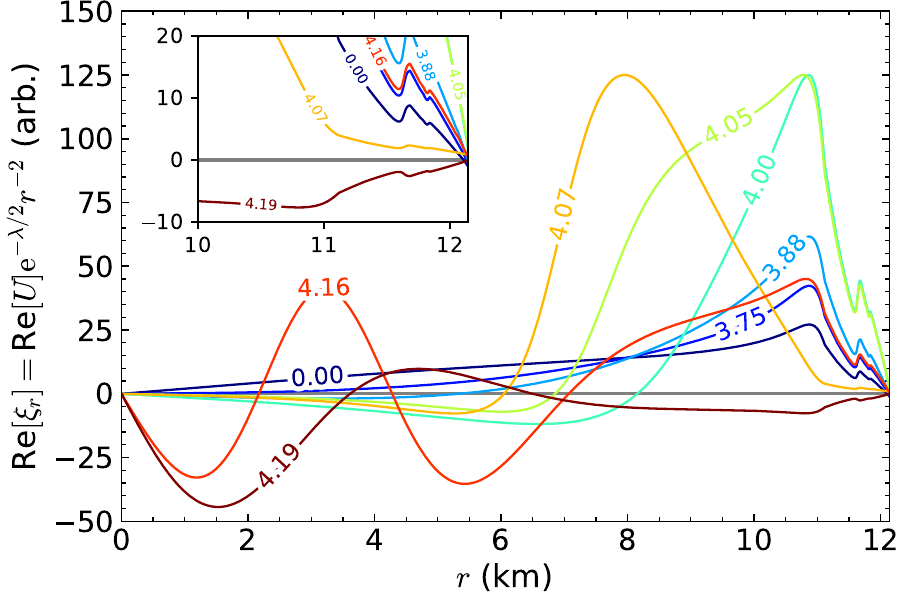}} \quad
\subfloat[QMC-RMF3, 
$g_2$-mode]{\includegraphics[width=0.38\linewidth]{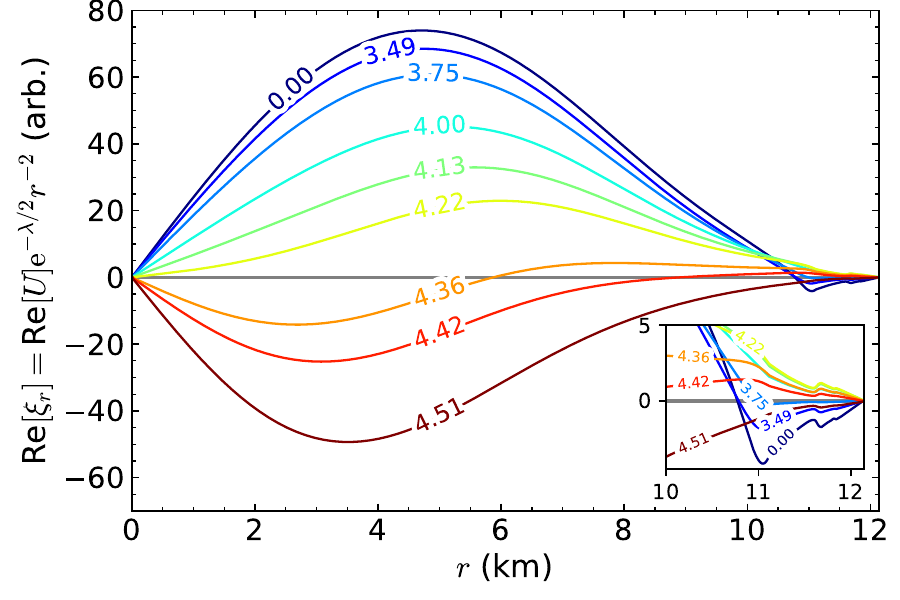}}\\
\subfloat[IUFSU, 
$g_1$-mode]{\includegraphics[width=0.38\linewidth]{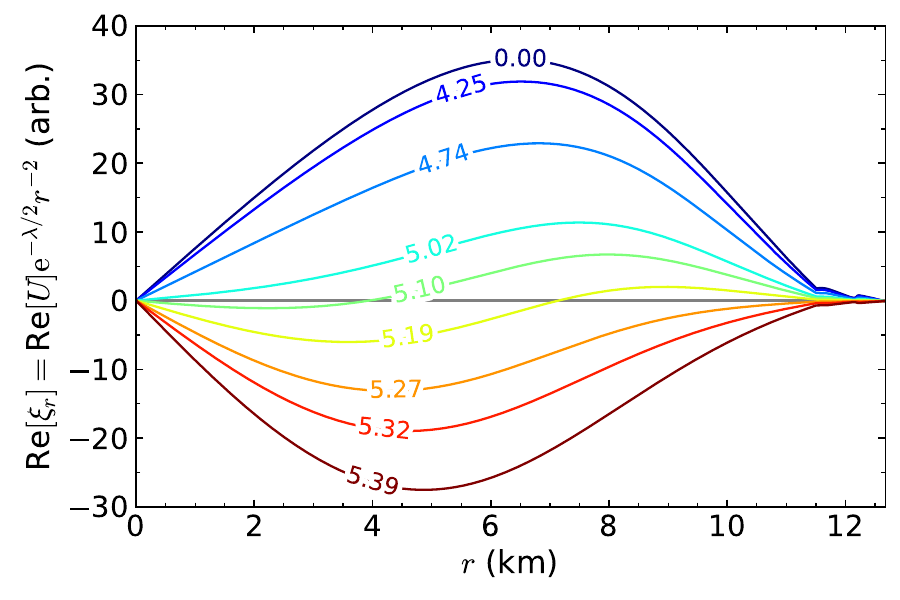}} \quad
\subfloat[IUFSU, 
$g_2$-mode]{\includegraphics[width=0.38\linewidth]{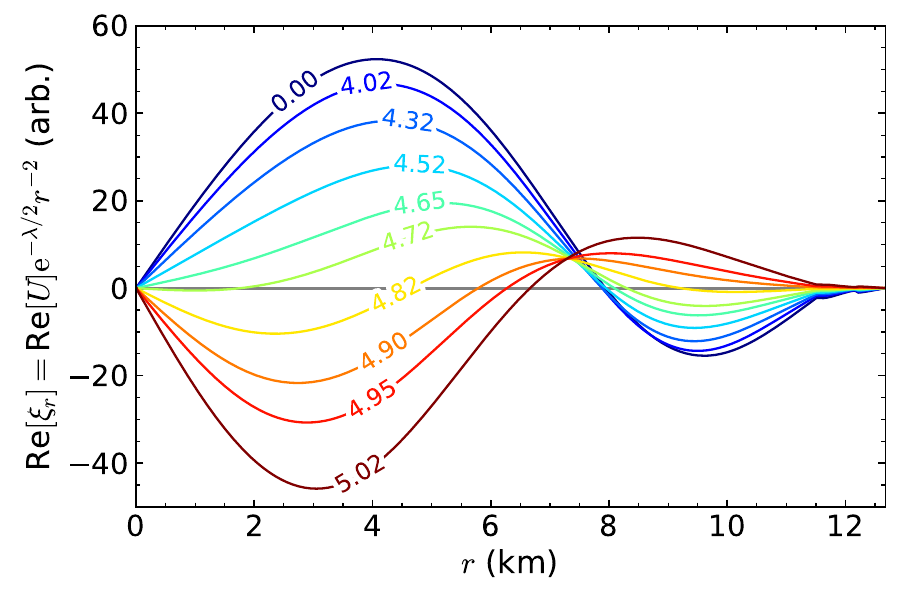}}\\
\subfloat[IOPB-I, 
$g_1$-mode]{\includegraphics[width=0.38\linewidth]{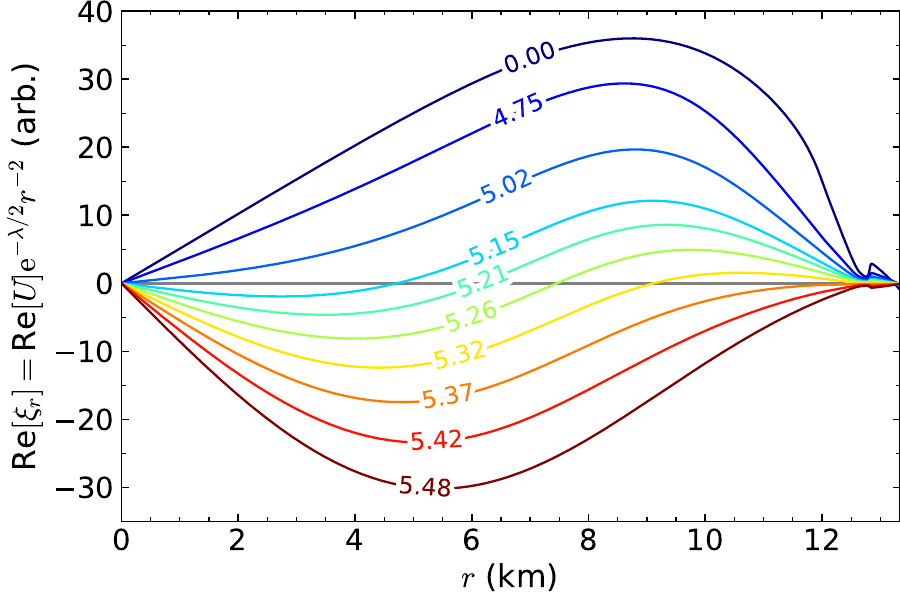}} \quad
\subfloat[IOPB-I, 
$g_2$-mode]{\includegraphics[width=0.38\linewidth]{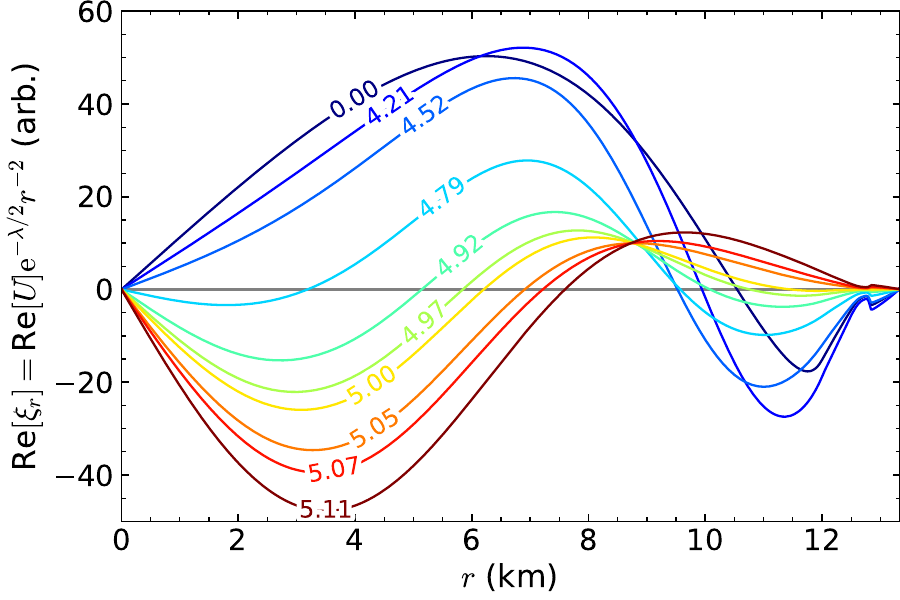}}
\caption{Real part of the radial displacement field \( \xi_r(r) \) for the fundamental 
\( g_1 \)-mode (left panels) and  first overtone 
\( g_2 \)-mode (right panels) as a function of the temperature, for three different EOSs: QMC-RMF3 (top row), IUFSU (middle row), and IOPB-I (bottom row). {The curves are labeled by the temperature in MeV. Zoomed-in insets show the displacement field at the outer edge of the star for both modes using the QMC-RMF3 EOS -- since many curves are overlapping in the insets, only some curves are labeled.} 
\label{fig:GModeDisplacementFields_allEOS}}
\end{figure*}

\begin{figure}
\centering
\includegraphics[width=0.9\linewidth]{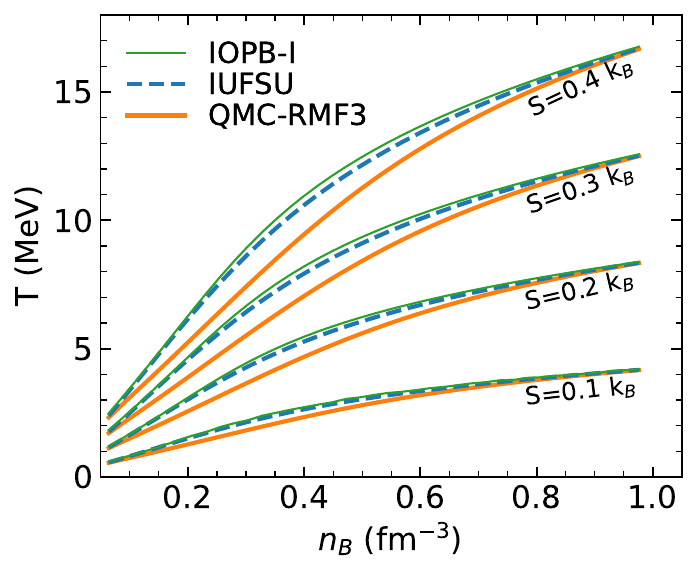}
\caption{Temperature profiles at fixed entropy per baryon for three (QMC-RMF3, IUFSU, IOPB-I) EOSs. These profiles are obtained by solving the equations of motion numerically in weak chemical equilibrium and at finite temperatures. At each density point, the entropy is calculated for several temperature points and then linearly interpolated in order to apply a root-finding algorithm to determine the desired constant entropy profile.}
\label{fig:T_profile_fix_S}
\end{figure}

We also calculated the \fmo of a 1.4 $\Msolar$ NS at various temperatures as shown in Fig.~\ref{fig:damping_time_f}, and found that the \fmo frequency is extremely insensitive to temperature. The real part of the frequency calculated with the Cowling approximation is 26\% larger than that with the full GR calculation, indicating that the Cowling approximation less accurate for studying $f$-mode oscillations~\citep{Sotani:2020mwc,Zhao:2022tcw,Rather:2024mtd}. 
The imaginary part is much smaller, so the damping time $\tau=1/{\rm Im}[\omega_f]$ 
is plotted instead. Since the Cowling approximation calculation ignores the damping due to gravitational waves completely, it results in a damping time $\approx 10$ seconds, much larger than the gravitational-wave damping time of the $f$-mode, $\approx 0.1$ seconds. Therefore, the damping introduced by bulk viscosity is much weaker than the damping from gravitational waves even at the resonant peak shown in Fig.~\ref{fig:contour_resonance}. 
This is not surprising given that the \fmo represents the compressible extension of the classical Kelvin mode, which corresponds to the fundamental surface gravity oscillation of an incompressible sphere~\citep{Kelvin1910}. In the incompressible limit, the \fmo displacement becomes divergence free with a radial dependence proportional to $\xi_r\propto r^l$, peaking near the stellar surface~\citep{chandrasekhar1960general}. Even in the compressible case, the \fmo maintains this qualitative structure with small divergence in Lagrangian fluid perturbation. As a result, the analytical solution for the Kelvin mode remains an excellent approximation to the \fmo  frequency, even in full GR~\citep{Zhao:2022tcw}. Therefore, the fluid undergoes compression and expansion only in the Eulerian description but remains largely uncompressed in the Lagrangian viewpoint. The smallness of Lagrangian compression is consistent with our finding that bulk viscosity has little impact on the linear \fmo in dense nucleonic matter across a wide range of temperatures. In contrast, the inclusion of additional degrees of freedom such as hyperons can significantly increase the bulk viscosity compared to the $npe$ estimate \Eqn{eq:bv_numerical} ~\citep{Ghosh:2023vrx,Alford:2020pld,Ofengeim:2019fjy}, reducing the damping time to $\lesssim 0.1$ s and allowing bulk viscosity to become the dominant dissipation mechanism.
However, the structure and frequency of the \fmo remain essentially unaffected. This contrasts with the case of $g$-modes, where large bulk viscosity and fast equilibration can affect both the real and imaginary parts of the frequency, potentially suppressing the mode entirely. 

When the bulk viscosity is included in the mode calculation, the resulting mode frequencies are no longer purely real numbers, so Sturm--Liouville theory does not strictly apply. This means that the displacement fields for the modes no longer follow the expected pattern whereby the number of nodes of the $(n-1)$th overtone $g$-mode is $n$, with the fundamental mode $g_1$ having 
one node, the first overtone having 
two nodes, etc. Instead, as the temperature increases and the bulk viscosity plays an increasingly important role, with the imaginary part of the mode frequency increasing, we observe that the number of 
nodes of the mode displacement field can change compared to the expectation from Sturm--Liouville theory. 

We demonstrate this in Fig.~\ref{fig:GModeDisplacementFields_allEOS} that shows the evolution in the displacement field for the fundamental and first overtone \gms as a function of the temperature for the QMC-RMF3 EOS. First, examining the fundamental mode in panel (a), one finds that for low temperatures $T\lesssim 3.5$ MeV, the displacement field takes an expected form for a fundamental mode with 
one node, but the displacement field below $r\sim6$ km moves closer to the Re$[\xi_r]=0$ axis as $T$ increases. As $T=4$ MeV is approached, the displacement field then crosses this axis close to $r=0$, with this node moving to higher $r$ as the temperature increases, while the node close to the outer boundary vanishes {as shown in the zoomed-in insets}, thus retaining a single {internal} node at first and then three 
nodes. At the highest temperatures examined, when the real part of the mode frequency becomes much smaller than the imaginary part, the displacement field has two 
nodes. The behavior of the first overtone mode displacement field in panel (b) also violates the typical behavior expected from Sturm--Liouville theory: as the temperature approaches $T=4$ MeV, the mode first becomes nodeless, with both its nodes near $r\approx 11$ km and the outer boundary vanishing {(this is difficult to distinguish even with the zoomed-in insets)}, then a new node develops at $r=0$, moving to larger $r$ as $T$ increases. The mode has one and then two nodes again before both vanish
at the highest temperature examined. The IUFSU and IOPB-I EOSs show similar behavior as seen in Fig.~\ref{fig:GModeDisplacementFields_allEOS} panels (c)--(f), with nodes appearing or disappearing as the temperature increases toward the limit where the imaginary part of $\omega_g$ exceeds the real part. 
For completeness, we note that the imaginary parts of the displacement fields are shown in in Appendix~\ref{app:RealImagSplit}. They exhibit similar radial structure to the real parts, with amplitudes much smaller at low temperatures and becoming comparable only once the imaginary part of the mode frequency becomes significant at temperatures $T \gtrsim 4$ MeV.

\begin{figure}
\centering
\includegraphics[width=0.85\linewidth]{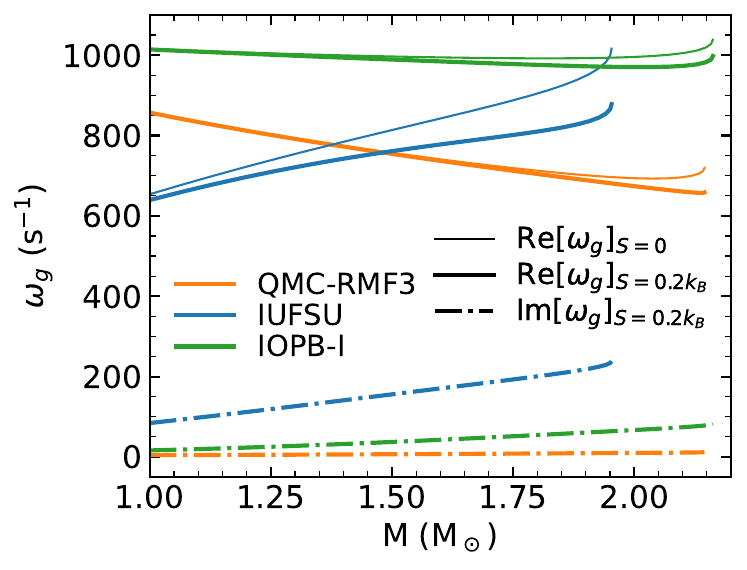}
\includegraphics[width=0.85\linewidth]{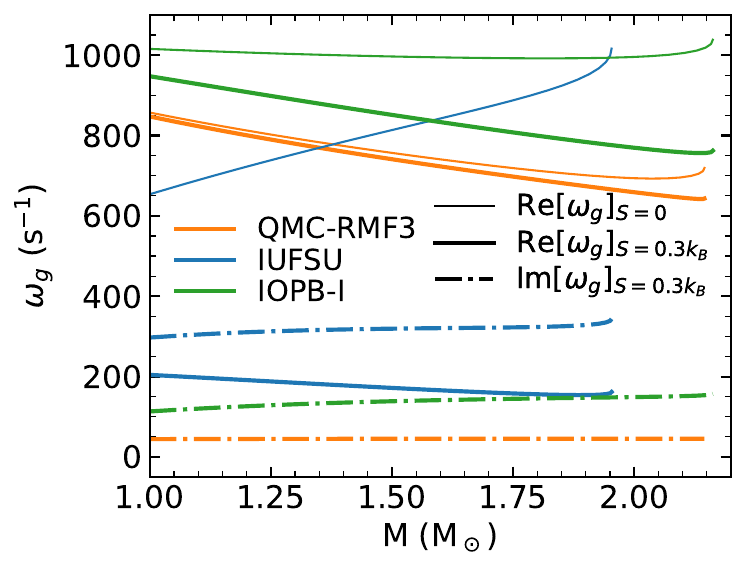}
\includegraphics[width=0.85\linewidth]{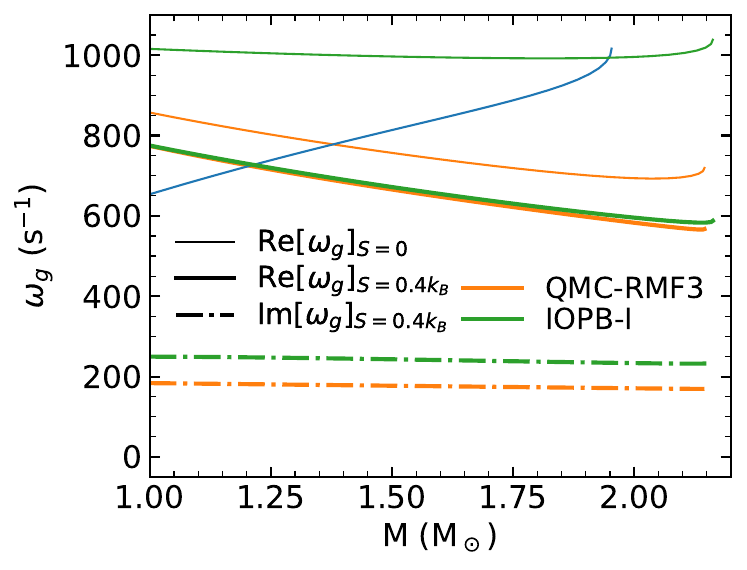}    
\caption{$g_1$-mode frequency as a function of NS mass at fixed entropy per baryon $S=0.2 \,k_B$ (top), $S=0.3 \,k_B$ (middle), and $S=0.4 \,k_B$ (bottom). The real (thick solid lines) and imaginary (dot-dashed lines) parts of the frequency are shown, and the thin solid line refer to the cold adiabatic \gm frequency in the absence of viscosity, which has no imaginary part. Results for three different EOSs are shown, and the calculations were performed within the Cowling approximation.}
\label{fig:omega_g_vs_M}
\end{figure}
\begin{figure}
\centering
\includegraphics[width=\linewidth]{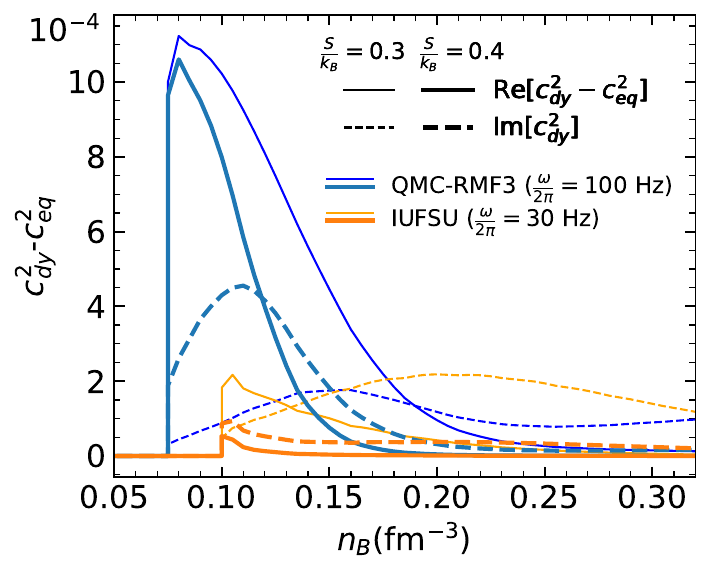}
\caption{Real (solid) and imaginary (dashed) parts of the difference between the dynamical and equilibrium sound speed squared, ${\rm Re}[\csqdy-\csqeq]$ and ${\rm Im}[\csqdy]$, as functions of the baryon number density $\nb$ at fixed values of the entropy per baryon, $S=0.3 \,k_B$ (thin) and $S=0.4 \,k_B$ (thick). Results are shown for the QMC-RMF3 EOS at an oscillation frequency $f=100$ Hz (blue) and for the IUFSU EOS at $f=30$ Hz (orange). The density regions where ${\rm Re}[\csqdy-\csqeq]>{\rm Im}[\csqdy]$ can support local \gm oscillations, whereas regions dominated by the imaginary part indicate strong damping and therefore the mode propagation is suppressed. 
}
\label{fig:FixedEntropySoundSpeedDiff}
\end{figure}
To investigate the impact of bulk viscosity on neutron stars with varying thermal structures~\citep{Kumar:2024jky,Ghosh:2023tbn}, we first compute temperature profiles at fixed entropy per baryon values, $S = s/\nb = \{ 0.1, 0.2, 0.3, 0.4\}\, k_B$, for three different EOSs as shown in Fig.~\ref{fig:T_profile_fix_S}. 
In the degenerate limit of a neutron gas, the entropy per baryon 
follows the relation $S = \pi^2\, k_B^2 T m_\mathrm{L}/{k_{\mathrm{F}}^2}$, 
where $k_{\mathrm{F}} = (3\pi^2\nb)^{1/3}$ is the Fermi momentum, and the Landau effective mass is defined as 
$m_\mathrm{L}=k_{\mathrm{F}} \left(dk_{\mathrm{F}}/d\eps_{\mathrm{F}} \right)$, which reduces to $m_\mathrm{L}=\sqrt{k_{\mathrm{F}}^2+m_D^2}$, where $m_D$ is the Dirac effective mass $m^*$ in the RMF model. Consequently, at fixed entropy per baryon, the temperature scales as $T \propto \nb^{2/3}/m_\mathrm{L}$, where 
the Landau effective mass 
$m_\mathrm{L}$ exhibits different density dependencies across the three EOSs. At subnuclear densities, where $k_\mathrm{F}\ll m_D$, the Landau mass approaches $m_\mathrm{L}\approx m_D$, yielding the scaling $T \propto \nb^{2/3}/m_D$. 
At several times nuclear saturation density, where  $k_{\mathrm{F}}\gg m_D$, one finds $m_\mathrm{L}\approx k_{\mathrm{F}}\propto \nb^{1/3}$, and the temperature scales as $T \propto \nb^{1/3}$. This trend of the power-law index decreasing from 2/3 to 1/3 appears in Fig.~\ref{fig:T_profile_fix_S}.

Using these temperature profiles, we compute the 
$g_1$-mode frequencies as a function of the stellar mass, shown in Fig.~\ref{fig:omega_g_vs_M}. For comparison, the zero entropy ($T=0$) 
$g_1$-modes, which are purely real, are also shown as thin lines. The most notable feature is that the real part of the frequency for the 
$g_1$-mode with the IUFSU EOS decreases much more with increasing $S$ than for the other two EOSs, such that it vanishes at $S=0.4\,k_B$. This can be explained by examining the temperature dependence of the sound speed squared difference $c_{\rm dy}^2-c_{\rm eq}^2$ for the IUFSU EOS and then comparing it to those obtained for the other two EOSs. At the crust-core transition density $\nb\approx 0.08$ fm$^{-3}$, the constant entropy $S=0.4\,k_B$ profile corresponds to a temperature $T\approx 3$ MeV according to Fig.~\ref{fig:T_profile_fix_S}. From Figs.~\ref{fig:cs2dy}--\ref{fig:cs2dy_dUrca}, for the QMC-RMF3 and IOPB-I EOSs, the real part of $c_{\rm dy}^2-c_{\rm eq}^2$ is larger than the imaginary part at this temperature and density. Though the density increases and temperature for fixed $S$ correspondingly increases deeper inside the star, the real part of $c_{\rm dy}^2-c_{\rm eq}^2$ being larger than the imaginary part near the crust-core transition is sufficient for the 
$g_1$-mode to be supported. However, for the IUFSU EOS, the real and imaginary parts of $c_{\rm dy}^2-c_{\rm eq}^2$ at $T\approx 3$ MeV are already nearly the same, and since the imaginary part only becomes greater and the real part smaller as the density increases and the temperature simultaneously increases for fixed $S$, the 
$g_1$-mode vanishes for this and larger values of $S$. This is shown explicitly in Fig.~\ref{fig:FixedEntropySoundSpeedDiff}, which compares the real and imaginary parts of the sound speed squared difference for $S=0.3\,k_B$ and $0.4\,k_B$ for the QMC-RMF3 and IUFSU EOSs as functions of the density, computed at values of $\omega$ approximating the 
$g_1$-mode frequency. 
For both values of $S$ for the QMC-RMF3 EOS, there is a region of the star where ${\rm Re}[c_{\rm dy}^2-c^2_{\rm eq}]>{\rm Im}[c_{\rm dy}^2]$, which is also the case for the IUFSU EOS when $S=0.3\,k_B$. 
However, ${\rm Im}[c_{\rm dy}^2]>{\rm Re}[c_{\rm dy}^2-c^2_{\rm eq}]$ for the IUFSU EOS at $S=0.4\,k_B$, indicating that there is no  
$g_1$-mode for this EOS at this value of $S$.

\section{Conclusions and outlook}
\label{sec:cons}

Thermal effects on the \gms of neutron stars are of increasing interest because these modes will be excited in protoneutron stars and post-merger remnant neutron stars. In this work, we introduce the concept of the dynamical sound speed squared, $\csqdy$, which captures the impact of bulk viscous effects in hot neutron star matter at temperatures of a few MeV. 
Unlike conventional adiabatic and equilibrium sound speeds, the dynamical sound speed incorporates the frequency-dependent damping effects of weak interaction equilibration, leading to a complex-valued expression. The real part of $\csqdy$ lies between the equilibrium sound speed squared $\csqeq$ and the adiabatic sound speed squared $\csqad$, while its imaginary part encodes dissipative effects associated with the bulk viscosity, $\zeta ={\rm Im}\left[\csqdy\right] {(\ep+P)}/{\omega}$. 
We demonstrate that the imaginary part ${\rm Im}[\csqdy]$ exhibits a resonant peak when the beta equilibration rate $\gamma$ becomes comparable to the oscillation frequency $\omega$. To further characterize this behavior, we identify the resonant region by mapping the peak values of ${\rm Im}[\csqdy]$ (across varying densities) in the $T-\omega$ plane and then compare it to the characteristic frequencies of neutron star oscillation modes. For EOSs where direct Urca is kinematically forbidden on the Fermi surface (e.g., QMC-RMF3), we find a single resonance band that starts at zero frequency around $T\approx 2$ MeV and extends to $\omega=100$ Hz at $T\approx 4.5$ MeV, continuing beyond. In contrast, EOSs with a direct Urca threshold exhibit significantly faster beta equilibration rates $\gamma$ at densities above the threshold, leading to an additional resonance band at lower temperatures or higher frequencies, as seen in the right panel of Fig.~\ref{fig:contour_resonance} for IOPB-I.

In order to assess the relative importance of bulk viscous damping and GW damping, we employ both full GR calculations and the relativistic Cowling approximation, in which metric perturbations are neglected. We validate that the Cowling approximation remains a reliable and accurate method for computing $g$-modes, even when bulk viscosity is included, but it is less accurate for the $f$-mode.

Using these methods, we investigate the effects of weak interactions and bulk viscosity on the \gms and \fmo of neutron stars by computing their complex frequencies at different temperatures. 
For $g$-modes, our results confirm that GW damping is negligible in comparison to bulk viscosity in warm neutron stars. As the temperature increases, bulk viscosity becomes more significant, leading to a decrease in the real part of the \gm frequency as well as an increase in its imaginary part, indicating stronger damping. We observe that for EOSs with a double-peak structure in the sound speed squared difference (e.g., QMC-RMF3), the odd and even overtones \gms behave differently, whereas EOSs with a single peak (e.g., IUFSU, IOPB-I) do not exhibit this distinction. Additionally, we find that at sufficiently high temperatures, bulk viscosity alters the nodal structure of the displacement field, deviating from the expectations of Sturm--Liouville theory, as modes develop additional nodes.

In contrast, we find that the \fmo frequency remains largely insensitive to temperature and that the bulk viscosity has a negligible effect on its damping compared to gravitational wave emission. This is consistent with the fact that the $f$-mode is 
the compressible extension of the incompressible Kelvin mode, which is nearly divergence-free. In the Lagrangian frame, fluid elements undergo minimal compression, significantly reducing the impact of bulk viscosity on the mode's dissipation.

The relative insensitivity of the \fmo to bulk viscosity is crucial for understanding the gravitational-wave spectrum from post-merger remnants. Since the \fmo emits gravitational waves efficiently, the gravitational-wave peak frequency is often dominated by \fmo oscillations~\citep{Ng:2020etb}, and in such cases, the bulk viscosity is expected to have only a minor impact on the gravitational-wave signal, as suggested in~\citet{Radice:2021jtw, Zappa:2022rpd}.
However, if the peak frequency contains nonlinear contributions from other modes, e.g. radial modes~\citep{Soultanis:2021oia}, then the bulk viscosity may substantially alter both the post-merger dynamics and the emitted gravitational waves~\citep{Most:2022yhe, Chabanov:2023blf}.

Our analysis shows that at temperatures of a few MeV, beta equilibration becomes rapid enough to significantly affect the frequency and damping time of $g$-modes. In addition to examining \gms at constant temperature, we also computed \gms for neutron stars with constant entropy per baryon profiles. This approach provides a more realistic representation of the thermal structures in newly formed or merging neutron stars. We find that increasing the entropy per baryon generally leads to a decrease in the real part of the \gm frequency, with the effect being most pronounced for the IUFSU EOS, where the frequency vanishes at $S=0.4 \,k_B$, due to strong bulk viscous effects. This behavior is directly linked to the temperature dependence of the sound speed squared difference $\csqdy-\csqeq$,  which determines the strength of the restoring force for 
$g$-modes. For EOSs like QMC-RMF3 and IOPB-I, where the real part of $\csqdy-\csqeq$ remains larger than the imaginary part at relevant densities, the \gm persists even at high entropy. However, for the IUFSU EOS, where the imaginary part dominates at all densities when the entropy per baryon reaches $S=0.4 \,k_B$, the restoring force weakens, leading to the disappearance of the 
$g_1$-mode. These results highlight that both the \gm frequency and its damping are highly sensitive to the thermal profile and the EOS when the temperature or entropy reaches the resonant peak of the bulk viscosity.

In our relativistic Cowling analysis, the $f$-, \gms and other uncalculated modes are treated as a linearly independent basis in the Hilbert space of fluid perturbations. 
Nonetheless, non-linear coupling between these modes may become significant in the highly dynamic environment of neutron star mergers. In such scenarios, the bulk viscous damping of \gms might indirectly influence the gravitational-wave signal via energy transfer or mode mixing with the dominant $f$-mode~\citep{Yu:2022fzw}. Besides, \gms play a critical role in the nonlinear saturation of unstable $f$-modes driven by the Chandrasekhar--Friedman--Schutz mechanism~\citep{Chandrasekhar1970,Friedman1978}. Specifically, coupling to stable \gms provides a dissipation channel that limits the \fmo amplitude~\citep{Pnigouras:2015bwa}. Our finding that bulk viscosity can strongly damp \gms at $T\gtrsim 5$ MeV suggests this saturation mechanism may be suppressed, potentially allowing the \fmo to grow to larger amplitudes than previously expected for cold NSs.

In the future, it is important to improve on the Fermi surface approximation of the Urca rates used in this work.  Finite-temperature effects blur the direct Urca threshold, eliminating the sharp jump in the beta equilibration rate that occurs at the threshold density~\citep{Alford:2018lhf,Alford:2021ogv}. This would modify the dynamic speed of sound and thus the $g$-modes.  In addition, a new method that consistently takes the in-medium collisions of the decaying nucleons into account, called the ``nucleon width approximation''~\citep{Alford:2024xfb}, would further improve the calculation of the beta equilibration rate $\gamma$.  On a different front, as the temperature rises above a few MeV, neutrino-trapping effects become important, and these should be taken into account. 

Matter at high densities is likely to include other degrees of freedom, including muons~\citep{Jaikumar:2021jbw,Alford:2021lpp}, pions~\citep{Kolomeitsev:2014gfa,Vijayan:2023qrt,Fore:2019wib,Harris:2024ssp,Pajkos:2024iry}, hyperons~\citep{Ofengeim:2019fjy,Alford:2020pld,Tran:2022dva,Li:2023owg,Kumar:2023ojk}, deconfined quarks~\citep{Alford:2019oge,Sotani:2023zkk,Constantinou:2023ged,Pradhan:2023zmg,Alford:2024tyj,Drago:2003wg}, and perhaps dark matter~\citep{Fornal:2023wji,Routaray:2022utr,Sen:2024yim,Shirke:2023ktu}.  These {exotic} particles have their own reaction channels that will contribute to chemical equilibration and will modify the bulk viscosity. Nevertheless, the dynamical sound speed introduced in this work provides a flexible framework that can incorporate such viscous effects in both radial and non-radial oscillation analyses~\citep{Zhang:2023zth}. 
{Additionally, the existence of a first-order phase transition between hadronic and quark matter in hybrid stars gives rise to a discontinuity $g$-mode.  Typically, this mode is calculated under the assumption that the quark-hadron conversion occurs arbitrarily slowly, similar to the historical development of compositional $g$-mode calculations~\citep{Miniutti:2002bh,Zhao:2022tcw,Rodriguez:2021sgk}.  However, quark-hadron conversion occurs at some finite, temperature-dependent rate, and we expect that accounting for this rate properly will alter the discontinuity $g$-mode frequency and damping time (see~\citet{Rau2023} for a treatment of the analogous radial mode problem). Of course, the discontinuity $g$-mode would vanish in the limit of fast quark-hadron conversion~\citep{Tonetto:2020bie}.}

Finally, to properly study the effects of weak interactions on $g$-modes in a physical context such as a core-collapse supernova or a binary NS merger, one must incorporate realistic, time-dependent thermal and density profiles from supernova and NS merger simulations~\citep{Ferrari2003,Camelio:2017nka,Sotani:2020mwc}. The thermal structure of a protoneutron star evolves dynamically, with significant changes in temperature, composition, and neutrino transport over short timescales. 
With these realistic profiles, one can study a new array of questions. For instance, in certain regions of a protoneutron star, where the local temperature and composition lead to thermal relaxation proceeding much faster than chemical equilibration, the conditions for the neutron finger instability may arise~\citep{Bruenn:2004dw}. 
Throughout this work, we assumed adiabatic compression, which excludes thermal relaxation and thereby keeps the oscillation stable against this instability. However, if the dynamical sound speed is defined at constant temperature rather than constant entropy, corresponding to the limit of very efficient thermal conduction, the onset of the neutron finger instability can be identified by examining the sign of the imaginary part of the dynamical sound speed.
Thus, using numerical supernova profiles will allow for a more accurate assessment of how bulk viscosity affects oscillation modes in astrophysical scenarios.

\begin{acknowledgments}
The authors thank Nils Andersson, Suprovo Ghosh, K.~J. Kwon, and David Radice for their valuable feedback on the manuscript.
We gratefully acknowledge the program ``Neutron Rich Matter on Heaven and Earth'' (INT-22r-2a), and the joint INT-N3AS workshop ``EOS Measurements with Next-Generation Gravitational-Wave Detectors'' (INT-24-89W), both held at the Institute for Nuclear Theory, University of Washington for hospitality and stimulating discussions. 
This research was supported in part by the INT's U.S. Department of Energy grant No. DE-FG02-00ER41132. 
T.Z. acknowledges support by the Network for Neutrinos, Nuclear Astrophysics and Symmetries (N3AS), through the National Science Foundation Physics Frontier Center, Grant No. PHY-2020275. 
P.B.R. was supported by the Simons Foundation through a SCEECS postdoctoral fellowship (grant No. PG013106-02). 
A.H.~acknowledges support by the U.S. Department of Energy, Office of Science, Office of Nuclear Physics, under Award No. DE-FG02-05ER41375. A.H.~furthermore acknowledges financial support by the UKRI under the Horizon Europe Guarantee project EP/Z000939/1. The work of S.P.H. was supported by the National Science Foundation grant PHY 21-16686. 
C.C. acknowledges support from the European Union's Horizon 2020 Research and Innovation Program under the Marie Sk\l{}odowska-Curie Grant Agreement No. 754496 (H2020-MSCA-COFUND-2016 FELLINI). 
The work of S.H. was supported by Startup Funds from the T.D. Lee Institute and Shanghai Jiao Tong University.

\end{acknowledgments}

\appendix

\section{Thermodynamic derivatives}
\label{app:thermodynamics}

In this section, we rewrite the thermodynamic derivatives in the main text to a basis of independent variables $\left\{x,\nb,T\right\}$ which are most convenient to calculate for a particular EOS. 
Recall, again, that $ S\equiv s/\nb$, $x=n_p/\nb$, $\delta x=x-x_{\textrm{eq}}$. Since equilibrium composition is determined as $x
_{\textrm{eq}}(\nb,S)$, $\partial \delta x$ is equivalent to $\partial x$. 
With the use of thermodynamic Jacobians, the derivatives in 
\Eqn{eq:gamma_def} 
can be written as 

\begin{eqnarray}
\dfrac{\partial \delta\mu}{\partial x}\bigg\vert_{\nb,S} &=& \dfrac{\partial \delta\mu}{\partial x}\bigg\vert_{\nb,T} - \dfrac{\dfrac{\partial \delta\mu}{\partial T}\bigg\vert_{x,\nb} \dfrac{\partial  S}{\partial x}\bigg\vert_{T,\nb} }{\dfrac{\partial  S}{\partial T}\bigg\vert_{x,\nb} },\label{eq:dmudx_nS}\\
\dfrac{\partial p}{\partial \nb}\bigg\vert_{x,S} &=& \dfrac{\partial p}{\partial \nb}\bigg\vert_{x,T}-\dfrac{\dfrac{\partial p}{\partial T}\bigg\vert_{\nb,x} \dfrac{\partial S}{\partial \nb}\bigg\vert_{x,T} }{\dfrac{\partial  S}{\partial T}\bigg\vert_{x,\nb} }, \label{eq:dpdn_xS}\\
\dfrac{\partial p}{\partial \nb}\bigg\vert_{\delta\mu,S}  &=&\dfrac{\partial p}{\partial \nb}\bigg\vert_{x,T}- \dfrac{A\dfrac{\partial p}{\partial x}\bigg\vert_{\nb,T}-B \dfrac{\partial p}{\partial T}\bigg\vert_{\nb,x} }{C}.\nonumber\\
\label{eq:dpdn_muS}
\end{eqnarray}
where
\begin{eqnarray}
A &=& \dfrac{\partial \delta\mu}{\partial \nb}\bigg\vert_{x,T}\dfrac{\partial  S}{\partial T}\bigg\vert_{\nb,x}-\dfrac{\partial \delta\mu}{\partial T}\bigg\vert_{\nb,x}\dfrac{\partial  S}{\partial \nb}\bigg\vert_{x,T} ,    \\
B &=& \dfrac{\partial \delta\mu}{\partial \nb}\bigg\vert_{x,T}\dfrac{\partial  S}{\partial x}\bigg\vert_{\nb,T} - \dfrac{\partial \delta\mu}{\partial x}\bigg\vert_{\nb,T}\dfrac{\partial  S}{\partial \nb}\bigg\vert_{x,T} ,   \\
C &=& \dfrac{\partial \delta\mu}{\partial x}\bigg\vert_{\nb,T}\dfrac{\partial  S}{\partial T}\bigg\vert_{x,\nb} - \dfrac{\partial \delta\mu}{\partial T}\bigg\vert_{x,\nb}\dfrac{\partial  S}{\partial x}\bigg\vert_{\nb,T}.
\end{eqnarray}
\begin{figure}
\centering
\includegraphics[width=\linewidth]{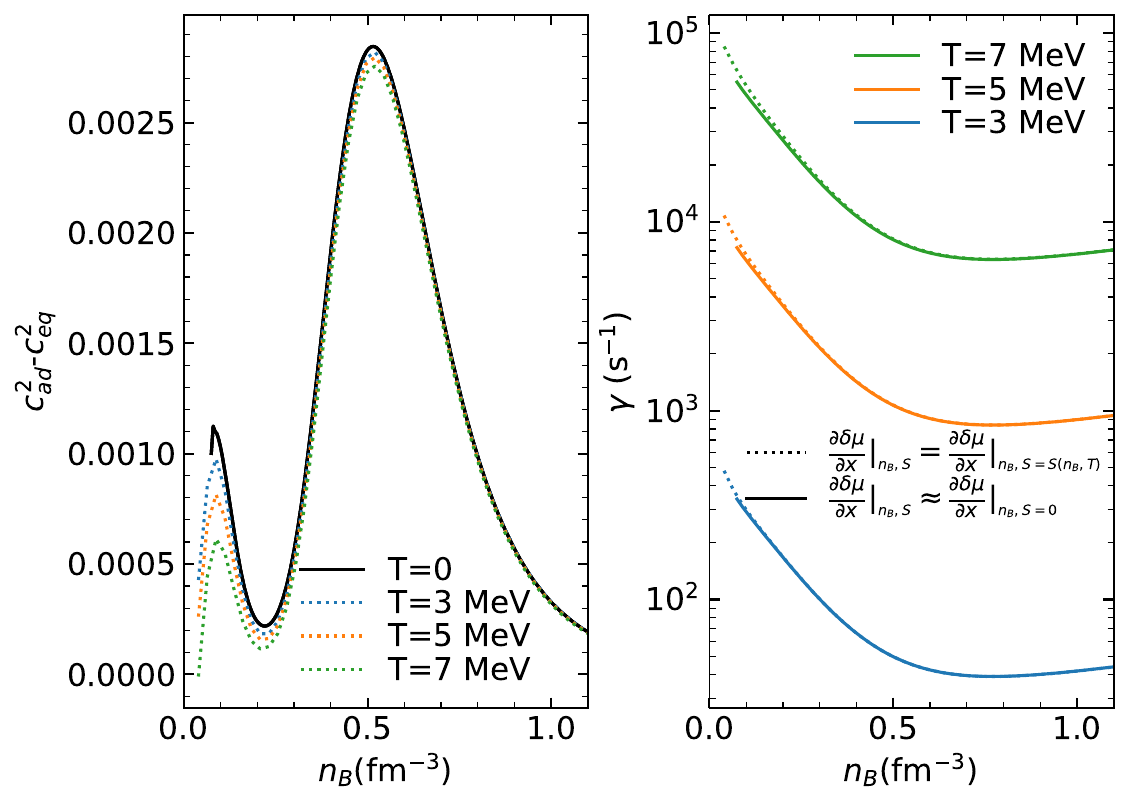}
\caption{Left panel: the differences between the adiabatic and equilibrium sound speed squared $\csqad-\csqeq$ around beta equilibrium $x_{\textrm{eq}}\,(\nb,T)$, as a function of the baryon density for various fixed-temperature profiles $T = [0,\,3,\,5,\,7]$ MeV {for the QMC-RMF3 EOS}. 
Right panel: the beta equilibration relaxation rate $\gamma$ as a function of $\nb$ for fixed-temperature profiles $T = [3,\,5,\,7]$ MeV. 
The dotted lines follow the exact definition in \Eqn{eq:gamma_def}.
The solid lines include finite-temperature effects in the beta reaction rate $\lambda_{\rm{n}\leftrightarrow\rm{p}}$, but approximate the susceptibility 
$\left. (\partial{\delta\mu}/\partial \delta x \right) |_{\nb,S}$ using its zero-temperature value. 
\label{fig:ignore_finiteT}}
\end{figure} 

The susceptibility (\ref{eq:dmudx_nS}) and the compressibilities (Eqs.~(\ref{eq:dpdn_xS}) and (\ref{eq:dpdn_muS}))
turn out to exhibit minimal temperature dependence in the NS core for temperatures $T \lesssim 10$ MeV. Fig.~\ref{fig:ignore_finiteT} illustrates the finite-temperature effects on these thermodynamic derivatives {for the QMC-RMF3 EOS}. The left panel shows the temperature-induced variation in the sound speed squared difference, $\csqad-\csqeq$, which depends on the compressibilities. This variation is small compared to its zero-temperature reference value (black solid line), and is even smaller relative to the sound speed squared themselves, particularly at high densities. Nevertheless, the difference at lower densities may have a more significant impact on crustal \gms\citep{Gittins:2024oeh}. 
The right panel displays the beta equilibration relaxation rate $\gamma$, which depends on the susceptibility and the net decay rate as defined in \Eqn{eq:net_decay_rate}. The solid lines represent the full computation of $\gamma$ from its definition in \Eqn{eq:gamma_def}, whereas the dashed lines correspond to calculations that neglect the temperature dependence of the susceptibility. The main finite-temperature contribution to $\gamma$ arises from the net decay rates rather than the susceptibility. 
Therefore, in this work, we neglect the temperature dependence of the susceptibility and compressibilities, 
and retain only the finite-temperature effects in the net decay rate. This approximation is employed in our calculations of both \fmo and $g$-modes.

\begin{figure}
\centering
\includegraphics[width=\linewidth]{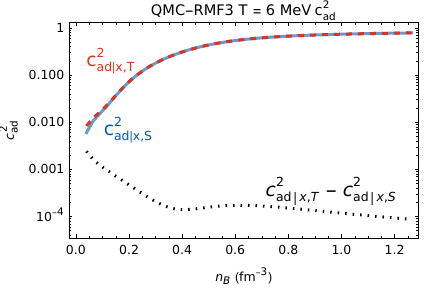}
\caption{{Isothermal (red, dashed) versus constant specific entropy (blue) adiabatic speed of sound $\csqad$ and their difference for the QMC-RMF3 EOS at a temperature of $T=6$ MeV. Even at temperatures at the upper range that we consider in this paper, the difference between the isothermal and the constant specific entropy $\csqad$ is negligible, especially at larger densities.}
\label{fig:comp_cs}}
\end{figure}

Figure~\ref{fig:comp_cs} shows a comparison of the adiabatic speed of sound from \Eqn{eq:cs2ad_cs2eq} calculated at constant temperature $T$ or constant specific entropy $S$. The relation between the required derivatives is shown in \Eqn{eq:dpdn_xS}. While we are calculating the speed of sound at constant $S$ in this work, the difference to the isothermal calculation is small, and only noticeable at densities below saturation densities, even at relatively high temperatures of $T=6$ MeV. 

\section{Real and imaginary parts splitting of mode equations}
\label{app:RealImagSplit}

To solve Eqs.~(\ref{eq:dUdr}) and (\ref{eq:dVdr}) by splitting into real and imaginary parts, we take $U=U_r+i\,U_i$, $V=V_r+i\,V_i$ and $\omega=\omega_r+i\,\omega_i$. We also write \Eqn{eq:GammaComplex} as \\ 

\begin{eqnarray}
\Gamma & = &  \Gamma_{\rm eq} \,+ \, \dfrac{\Gamma_{\rm ad}-\Gamma_{\rm eq}}{1-\dfrac{\gamma}{\omega}i} 
\nonumber\\
&=&\Gamma_{\rm eq}\,+ \, \dfrac{(\Gamma_{\rm ad}-\Gamma_{\rm eq})\,(\omega_r^2+\omega_i(\omega_i-\gamma))}{\omega_r^2 \, + \, (\omega_i-\gamma)^2} \nonumber \\
&& + \, i \, \dfrac{(\Gamma_{\rm ad}-\Gamma_{\rm eq})\,\omega_r\gamma}{\omega_r^2 \, + \, (\omega_i-\gamma)^2}
\nonumber\\
&\equiv& \Gamma_r\, + \, i\Gamma_i,
\label{eq:GammaSplit}
\end{eqnarray}
where $\gamma$ is defined in \Eqn{eq:gamma_def}. 
Also defining the real and imaginary parts of $N^2$ as
\begin{align}
N^2={}&g^2\dfrac{\ep+p}{p}\left(\dfrac{1}{\Gamma_{\rm eq}}-\dfrac{\Gamma_r}{\Gamma_r^2+\Gamma_i^2}\right){\rm e}^{\nu-\lambda}
\nonumber\\
{}&+ig^2\dfrac{\ep+p}{p}\left(\dfrac{\Gamma_i}{\Gamma_r^2+\Gamma_i^2}\right){\rm e}^{\nu-\lambda}\equiv N_r+iN_i,
\end{align}
the four mode equations we solve are
\begin{widetext}
\begin{align}
\dfrac{{\rm d}U_r}{{\rm d}r}={}&\dfrac{(\ep+p)g}{p(\Gamma_r^2+\Gamma_i^2)}\left(\Gamma_rU_r+\Gamma_iU_i\right)
+{\rm e}^{\lambda/2}\left(\dfrac{l(l+1){\rm e}^{\nu}(\omega_r^2-\omega_i^2)}{(\omega_r^2+\omega_i^2)^2}-\dfrac{(\ep+p)r^2\Gamma_r}{p(\Gamma_r^2+\Gamma_i^2)}\right)V_r
\nonumber
\\
{}&+{\rm e}^{\lambda/2}\left(\dfrac{l(l+1){\rm e}^{\nu}2\omega_r\omega_i}{(\omega_r^2+\omega_i^2)^2}-\dfrac{(\ep+p)r^2\Gamma_i}{p(\Gamma_r^2+\Gamma_i^2)}\right)V_i,
\\
\dfrac{{\rm d}U_i}{{\rm d}r}={}&\dfrac{(\ep+p)g}{p(\Gamma_r^2+\Gamma_i^2)}\left(-\Gamma_iU_r+\Gamma_rU_i\right)
+{\rm e}^{\lambda/2}\left(-\dfrac{l(l+1){\rm e}^{\nu}2\omega_r\omega_i}{(\omega_r^2+\omega_i^2)^2}+\dfrac{(\ep+p)r^2\Gamma_i}{p(\Gamma_r^2+\Gamma_i^2)}\right)V_r
\nonumber
\\
{}&+{\rm e}^{\lambda/2}\left(\dfrac{l(l+1){\rm e}^{\nu}(\omega_r^2-\omega_i^2)}{(\omega_r^2+\omega_i^2)^2}-\dfrac{(\ep+p)r^2\Gamma_r}{p(\Gamma_r^2+\Gamma_i^2)}\right)V_i,
\\
\dfrac{{\rm d}V_r}{{\rm d}r}={}&\dfrac{{\rm e}^{\lambda/2-\nu}}{r^2}\left[(\omega_r^2-\omega_i^2-N_r)U_r-(2\omega_r\omega_i-N_i)U_i\right]
+\dfrac{(\ep+p)g}{p}\left[\left(\dfrac{1}{\Gamma_{\rm eq}}-\dfrac{\Gamma_r}{\Gamma_r^2+\Gamma_i^2}\right)V_r-\dfrac{\Gamma_i}{\Gamma_r^2+\Gamma_i^2}V_i\right],
\\
\dfrac{{\rm d}V_i}{{\rm d}r}={}&\dfrac{{\rm e}^{\lambda/2-\nu}}{r^2}\left[(2\omega_r\omega_i-N_i)U_r+(\omega_r^2-\omega_i^2-N_r)U_i\right]
+\dfrac{(\ep+p)g}{p}\left[\dfrac{\Gamma_i}{\Gamma_r^2+\Gamma_i^2}V_r+\left(\dfrac{1}{\Gamma_{\rm eq}}-\dfrac{\Gamma_r}{\Gamma_r^2+\Gamma_i^2}\right)V_i\right].
\end{align}
\end{widetext}

The boundary conditions for the complex $U$ and $V$ cases are simply the real and imaginary parts of the real $U$ and $V$ boundary conditions. At $r=0$ we have
\begin{align}
U_r(r=0)=&\dfrac{l(\omega_r^2-\omega_i^2)}{(\omega_r^2+\omega_i^2)^2}Y_0r^{l+1},
\\
U_i(r=0)=&-\dfrac{2l\omega_r\omega_i}{(\omega_r^2+\omega_i^2)^2}Y_0r^{l+1},
\\
V_r(r=0)=&\dfrac{\ep}{\ep+p}Y_0r^{l},
\\
V_i(r=0)=&0,
\end{align}
and at $r=R$ we have
\begin{align}
0=V_r(r=R)-\dfrac{1}{2R^2}U_r(r=R){\rm e}^{-\lambda(r=R)/2}\left.\dfrac{{\rm d}\nu}{{\rm d}r}\right|_{r=R},
\\
0=V_i(r=R)-\dfrac{1}{2R^2}U_i(r=R){\rm e}^{-\lambda(r=R)/2}\left.\dfrac{{\rm d}\nu}{{\rm d}r}\right|_{r=R}.
\label{eq:OuterBCSplit}
\end{align}

\begin{figure*}
\centering
\subfloat[QMC-RMF3, 
$g_1$-mode]{\includegraphics[width=0.38\linewidth]{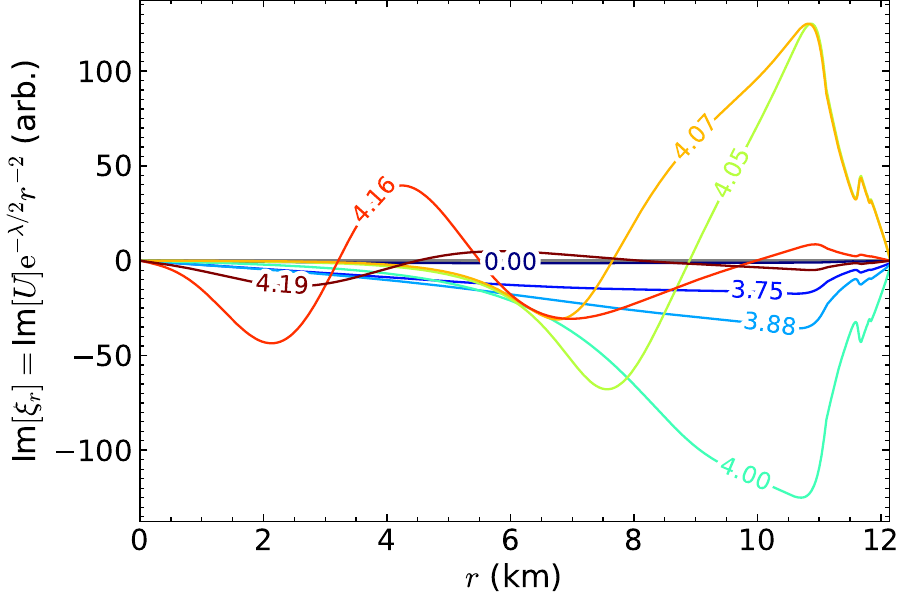}} \quad
\subfloat[QMC-RMF3, 
$g_2$-mode]{\includegraphics[width=0.38\linewidth]{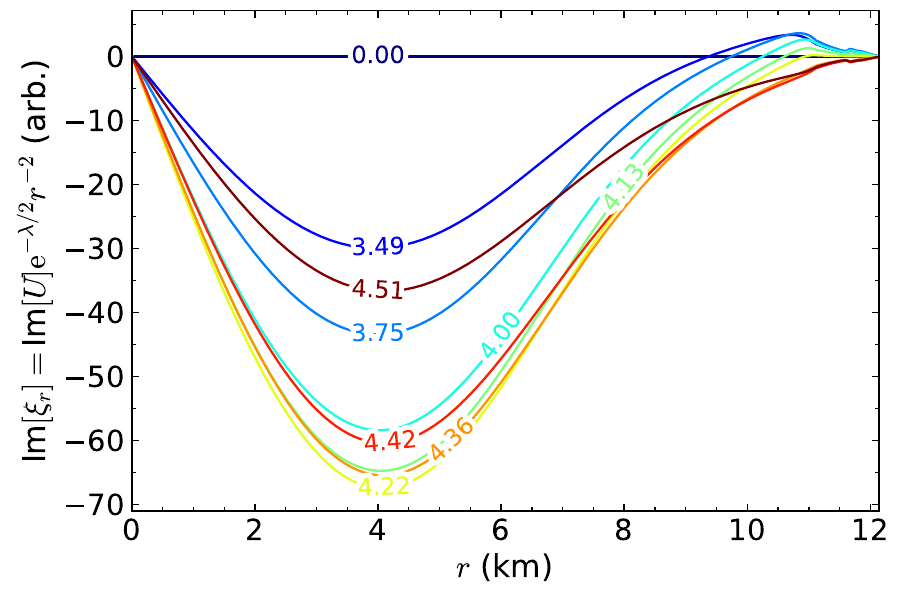}}\\
\subfloat[IUFSU, 
$g_1$-mode]
{\includegraphics[width=0.38\linewidth]{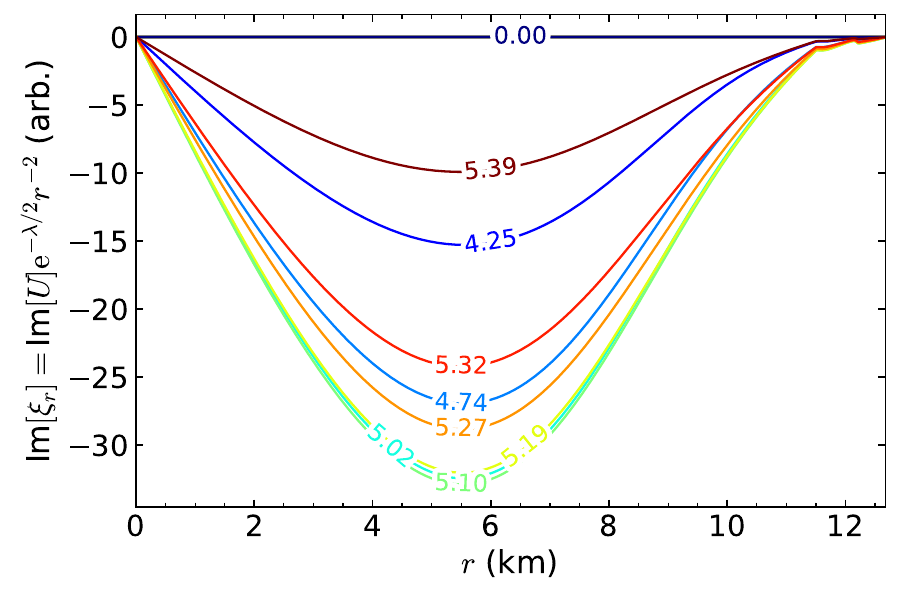}} \quad
\subfloat[IUFSU, 
$g_2$-mode]
{\includegraphics[width=0.38\linewidth]{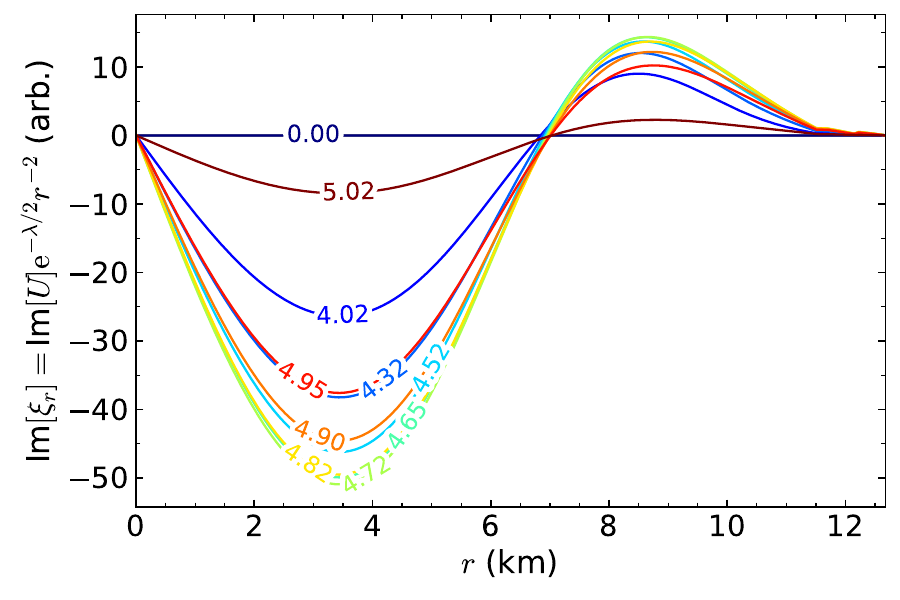}}\\
\subfloat[IOPB-I, 
$g_1$-mode]
{\includegraphics[width=0.38\linewidth]{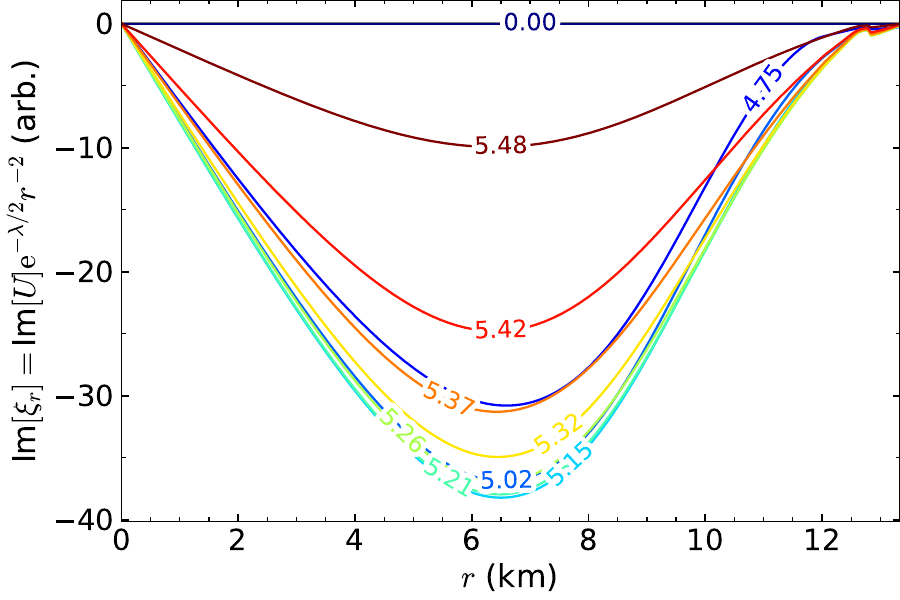}} \quad
\subfloat[IOPB-I, 
$g_2$-mode]{\includegraphics[width=0.38\linewidth]{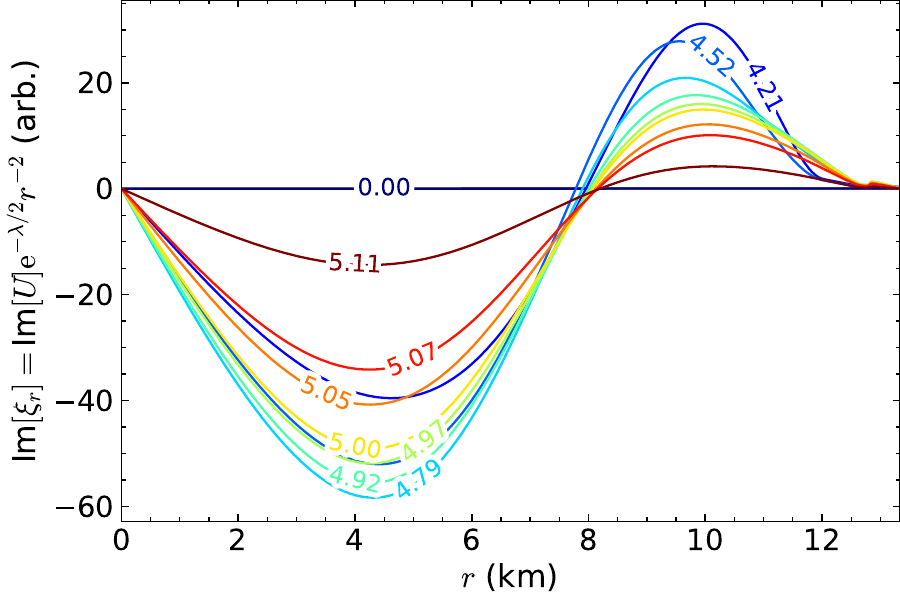}}
\caption{Imaginary part of the radial displacement field \( \xi_r(r) \) for the fundamental 
\( g_1 \)-mode (left panels) and  first overtone 
\( g_2 \)-mode (right panels) as a function of the temperature, for three different EOSs: QMC-RMF3 (top row), IUFSU (middle row), and IOPB-I (bottom row).}\label{fig:GModeDisplacementFieldsImag_allEOS}
\end{figure*}

Figure~\ref{fig:GModeDisplacementFields_allEOS} in the main text shows the real part of the radial displacement field, $U_r$.
For completeness, we display in Fig.~\ref{fig:GModeDisplacementFieldsImag_allEOS} 
the corresponding imaginary parts for the $g_1$-mode and $g_2$-mode, 
calculated for the QMC-RMF3, IUFSU, and IOPB-I EOSs. The imaginary parts exhibit radial structures similar to those of the real parts. 
At low temperatures their amplitudes are much smaller, but they grow and become comparable 
to the real parts once the imaginary part of the mode frequency is significant 
($T \gtrsim 4$ MeV).


\bibliography{paper}
\end{document}